\documentclass[10pt,letterpaper]{article}
\usepackage[top=0.85in,left=2.75in,footskip=0.75in]{geometry}

\usepackage[utf8]{inputenc} 
\usepackage[T1]{fontenc}    
\usepackage{hyperref}       
\usepackage{url}            
\usepackage{booktabs}       
\usepackage{amsfonts}       
\usepackage{nicefrac}       
\usepackage{microtype}      
\usepackage{graphicx}
\usepackage{tabularx}
\usepackage{multirow}
\usepackage{algorithm}
\usepackage{algpseudocode}
\usepackage{amsmath}
\usepackage{hyperref}
\usepackage{float, subfig}
\usepackage{makecell}
\usepackage{xcolor}
\usepackage{soul}
\usepackage[right]{lineno}
\newcolumntype{+}{!{\vrule width 2pt}}

\newlength\savedwidth

\raggedright
\usepackage[aboveskip=1pt,labelfont=bf,labelsep=period,justification=raggedright,singlelinecheck=off]{caption}

\makeatletter
\renewcommand{\@biblabel}[1]{\quad#1.}
\makeatother

\usepackage{lastpage,fancyhdr,graphicx}
\usepackage{epstopdf}
\fancyheadoffset[L]{2.25in}
\fancyfootoffset[L]{2.25in}
\begin{document}
\vspace*{0.2in}

\begin{flushleft}
{\Large
\textbf\newline{{Detecting COVID-19 from digitized ECG printouts using 1D convolutional neural networks} \\
}}
\bigskip
\textbf{Thao Nguyen\textsuperscript{1},} 
\textbf{Hieu H. Pham\textsuperscript{1,2},}
\textbf{Khiem H. Le\textsuperscript{1},}
\textbf{Anh-Tu Nguyen\textsuperscript{1},}
\textbf{Tien Thanh\textsuperscript{3},}
\textbf{Cuong Do\textsuperscript{1,*}}
\\
\bigskip
\textbf{1} College of Engineering \& Computer Science, VinUniversity, Hanoi, Vietnam
\\
\textbf{2} VinUni-Illinois Smart Health Center, VinUniversity, Hanoi, Vietnam\\
\textbf{3} Pulmonology, Vinmec International Hospital, Hanoi, Vietnam
\\
\bigskip

%
%





* Corresponding author: cuong.dd@vinuni.edu.vn

\end{flushleft}
\section*{Abstract}

The COVID-19 pandemic has exposed the vulnerability of healthcare services worldwide, raising the need to develop novel tools to provide rapid and cost-effective screening and diagnosis. Clinical reports indicated that COVID-19 infection may cause cardiac injury, and electrocardiograms (ECG) may serve as a diagnostic biomarker for COVID-19. This study aims to utilize ECG signals to detect COVID-19 automatically. We propose a novel method to extract ECG signals from ECG paper records, which are then fed into one-dimensional convolution neural network (1D-CNN) to learn and diagnose the disease. To evaluate the quality of digitized signals, R peaks in the paper-based ECG images are labeled. Afterward, RR intervals calculated from each image are compared to RR intervals of the corresponding digitized signal. Experiments on the COVID-19 ECG images dataset demonstrate that the proposed digitization method is able to capture correctly the original signals, with a mean absolute error of 28.11 ms. The 1D-CNN model (SEResNet18), which is trained on the digitized ECG signals, allows to identify between individuals with COVID-19 and other subjects accurately, with classification accuracies of 98.42\% and 98.50\% for classifying COVID-19 \textit{vs.} Normal and COVID-19 \textit{vs.} other classes, respectively. Furthermore, the proposed method also achieves a high-level of performance for the multi-classification task. Our findings indicate that a deep learning system trained on digitized ECG signals can serve as a potential tool for diagnosing COVID-19. 



\section*{Introduction}
    Since the initial infection was reported in Wuhan, China at the end of 2019, COVID-19 has become a global pandemic, infecting more than 500 million individuals and causing more than 6 million deaths (until the end of April, 2022) ~\cite{diseases2022transitioning}. 
    COVID-19 is becoming an endemic disease that will always be with us~\cite{diseases2022transitioning,antia2021transition}. COVID-19 disease, when it becomes an endemic disease, may still pose problems for many people, especially vulnerable groups like the elderly and those with pre-existing medical conditions, owing to its rapid contagious~\cite{mao2020implications}. As a result, detecting COVID-19 and developing an effective treatment strategy for patients require a fast and accurate diagnosis.
    
    To diagnose and screen COVID-19 infections, previous studies used a combination of genome sequencing, nucleic acid molecular testing, clustered regularly interspaced short palindromic repeats editing techniques, antigen/antibody detection, and computed tomography imaging. Because of their excellent sensitivity and specificity, polymerase chain reaction (PCR)-based assays are considered the gold standard for virus identification, and Reverse Transcription PCR (RT-PCR) is the most commonly used diagnostic method~\cite{shen2020recent}. However, PCR-based detection has some drawbacks, including the need for a high-purity sample, expensive laboratory equipment, expert training, and a long response time~\cite{corman2020detection}. Other molecular diagnostic techniques, such as clustered regularly interspaced short palindromic repeats (CRISPR) and gene sequencing, suffer from the same limitations. Chest CT detection is also a significant method to diagnose patients suspected of having a COVID-19 infection. Most COVID-19 patients have characteristic radiologic findings in chest CT scans, such as multifocal plaque consolidation and ground glass opacity, according to several studies~\cite{chung2020ct,li2020ct}. However, Bernheim \textit{et al.}~\cite{bernheim2020chest}  discovered that CT scans are unable to diagnose 56\% of patients in the early stages of symptom development. Some studies have also reported that some patients who tested positive for RT-PCR show initially normal X-rays or chest CT scans due to the lung may not be the target organ of the SARS-CoV-2 infection~\cite{bernheim2020chest, ai2020correlation}.
    
    
    Previous studies have shown that COVID-19 may be responsible for various cardiovascular complications, which has been the impetus for studies using ECG signals to diagnose COVID-19~\cite{soumya2021impact}. The ECG is an excellent indication of pathological abnormalities in the cardiovascular system. 
    This research is aimed at the possibility of diagnosing COVID-19 using paper-based ECG data and Deep Learning (DL) techniques, taking advantage of the most promising ECG qualities including accessibility, dependability, cheap cost, real-time monitoring, and harmlessness. The technique presented in this research is expected to be an accurate and effective alternative to fill the COVID-19 detection gap.
    
    Several previous studies have also been conducted to evaluate the feasibility of using ECG signals as a diagnostic method for COVID-19. Khan \textit{et al.}~\cite{khan2021ecg} built a dataset of 1937 ECG images of Cardiac and COVID-19 patients, which covered 5 classes: COVID-19, Abnormal Heartbeat, Myocardial Infarction (MI), Previous History of MI (RMI), and Normal Person. This dataset facilitates subsequent studies to investigate the ability to detect COVID-19 based on ECG signals. 
    Ozdemir \textit{et al.} proposed a method called hexaxial feature mapping to represent 12-lead ECG to 2D colorful images~\cite{ozdemir2021classification}. Specifically, Gray-Level Co-Occurrence Matrix (GLCM) method was used to extract features and generate hexaxial mapping images. These generated images are then fed into a new CNN architecture to diagnose COVID-19. 
    Irmak in~\cite{irmak2022covid} introduced a novel deep CNN model to demonstrate the feasibility of using ECG signals to diagnose COVID-19 and yielded an overall classification accuracy of 98.57\%, 93.20\%, 96.74\% for COVID-19 vs. Normal, COVID-19 vs. Abnormal Heartbeats and COVID-19 vs. MI binary classification tasks, respectively.
    Attallah in~\cite{attallah2022ecg} proposed a pipeline called ECG-BiCoNet, which makes use of five different deep learning models with different structural designs. 
    Nevertheless, most of the abovementioned studies utilized ECG images as direct input into machine learning algorithms. These approaches do not exploit the advantages of 1D signals such as high temporal resolution, saving computational costs and storage space. Furthermore, feeding images directly into a deep learning model makes the model sensitive to the type of ECG report form (various ECG printout templates available depending on the type of ECG machine and ECG file reading program) and image resolution.
    
    In this study, a novel image-to-signal conversion algorithm is proposed, which extracts 1D ECG signal from ECG printout images. This algorithm is then tested in COVID-19 classification application to evaluate its efficiency and accuracy. To the best of our knowledge, this is the first time the ECG signal extracted from ECG printouts was used to detect COVID-19. This method allows us to normalize data from many sources, removes the model's reliance on image quality, and enhances computing performance by using 1D convolution. 
    
    In addition, the image-to-signal conversion also benefits ECG data storage. The majority of ECG records were collected in paper-based form over the years~\cite{sun2019novel}. However, it has the drawbacks of being damaged over time and being difficult to archive data in order to establish an electronic medical record database. A good storage effort may be a viable answer to these issues. The digitization procedure is one of the most efficient ways to save ECG data. The ECG printouts can be digitized to reduce storage space and to convey various vital morphological and functional information for clinical and diagnostic scientific study. This digital data can be then analyzed with a machine learning algorithm to detect heart abnormalities quickly. Previous studies have developed algorithms to convert paper data into 1D signals~\cite{sun2019novel,mishra2021ecg,li2020deep,tabassum2020numerical}. These algorithms, however, were either too complicated or inadequately conceived for our experimental data. In this study, we developed a simple, rapid, and effective method, based on a pixel tracking algorithm for converting an ECG image to signals. After being extracted, the 12 signal segments from 12 leads will be concatenated and fed into the SEResNet18 model~\cite{hu2018squeeze}, which will classify ECG signals of individuals with COVID-19 and other CVDs. Several scenarios were undertaken to demonstrate the capability of identifying COVID-19 based on ECG signals. In addition, we evaluated our approach by comparing it against previous studies. The findings demonstrated that our approach performs competitively and ECG could be a useful diagnostic tool for COVID-19.
    
    In summary, our main contributions are threefold:
    \begin{itemize}
        
        \item First, we develop a high-precision COVID-19 vs. other CVDs vs. healthy individuals classification system that takes the ECG image as direct input, then the ECG signal is extracted and fed into a 1D-CNN.
        
        \item Second, a novel algorithm for converting ECG paper records to digital signals was developed, which took ECG images as input and converted them into electronic ECG recording forms.
        
        \item Third, extracted 1D ECG signals are used as input for the machine learning model in the COVID-19 diagnosis task for the first time, to the best of our knowledge. Extracting ECG signals from images provides the following advantages: (i) computational cost savings due to the use of 1D-CNN; (ii) application in converting ECG paper into soft files for archiving; (iii) hand-crafted feature extraction methods can be applied to ECG signals for further analysis.
    
        \item Last, by simulating the data settings of prior research and conducting extensive experiments, we demonstrate that models using ECG signal as input outperform the current state-of-the-art, which does the classification directly on the ECG image. Our codes for signal-to-image conversion and modeling are made available at \url{https://github.com/vinuni/DiECG-Cov}.
    \end{itemize}
    
    This paper is structured as following. In \nameref{section:methods} section, the image-to-signal conversion method is explained. In section~\nameref{section:experiments}, the used dataset is presented first, followed by a description of the experimental settings and evaluation criteria that we employ to assess model performance.In \nameref{section:results} section, the experimental results are presented and analyzed. The main findings and limitations of the study are summarized and some useful suggestions are provided in \nameref{section: discussions}. Finally, the entire study are summarized in the \nameref{section:conclusion} section.

\section*{Methods}
\label{section:methods}
    
\begin{figure}[hbt!]
      \centering
      \includegraphics[width=0.8\linewidth, scale=0.8]{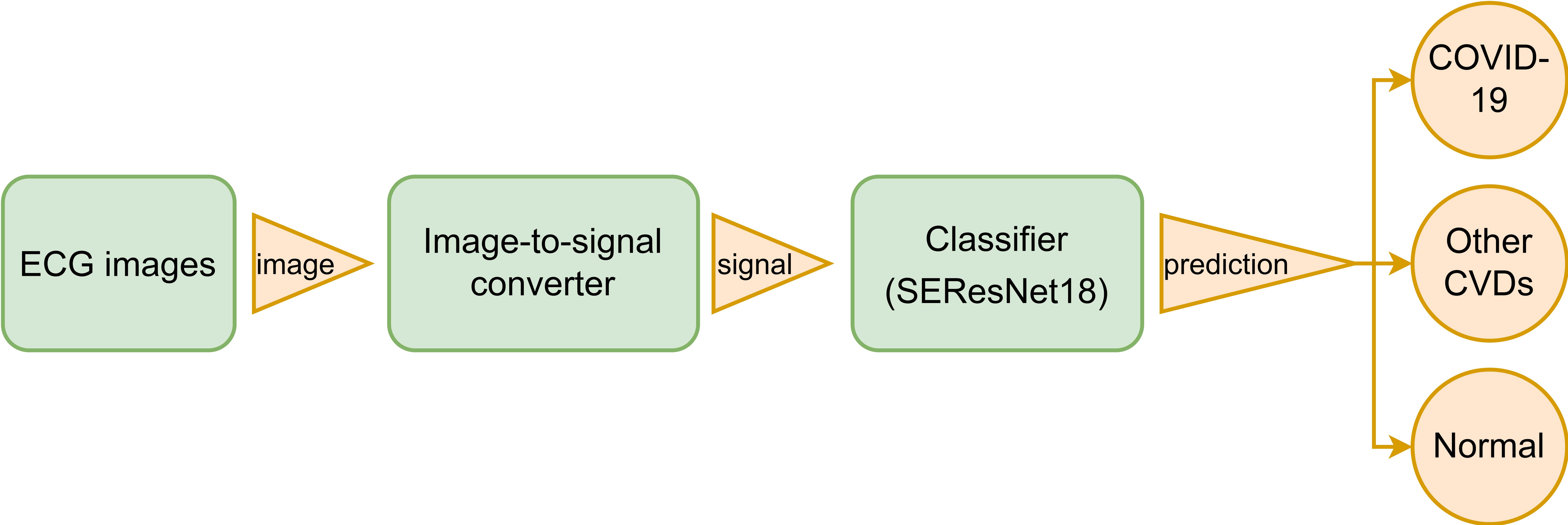}
      \caption{{\bf Schematic diagram of COVID-19 identification system. The process consists of 2 steps: converting ECG images to signals and classifying the resulting ECG signals.}}
      \label{fig:diagram}
    \end{figure}

    Our proposed approach for identifying COVID-19 from ECG images involves two steps: converting images into 1D signals and feeding these signals into a CNN model that distinguishes between COVID-19, other CVDs, and normal ECG. The schematic diagram for the proposed system is illustrated in Fig~\ref{fig:diagram}.
    
    \subsection*{Image-to-signal conversion algorithm}
    
\begin{figure}[hbt!]
      \centering
      \includegraphics[width=0.98\linewidth, scale=0.8]{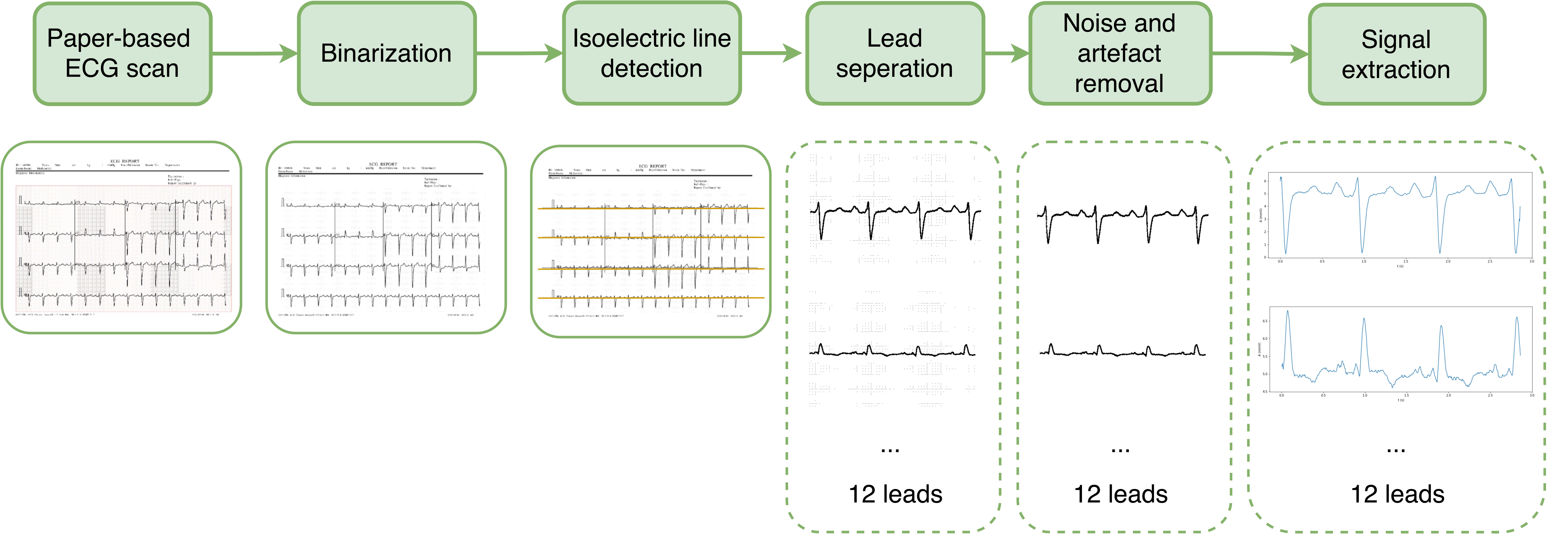}
      \caption{{\bf Procedure for converting paper-based ECG scans to 1D signals.} The process of converting image to signal consists of 5 steps: image binarization, isoelectric line detection, lead separation, noise removal and signal extraction.}
      \label{fig:Fig2}
    \end{figure}

    ECG data is often stored in paper form, which consumes storage space, and as a machine learning model input, 2D image data requires more computational cost than time series data (1D signal). For those reasons, we extract the 1D ECG signal from the scanned ECG paper and use that signal as input to the classifier. The procedure for extracting the ECG signal consists of 5 steps as depicted in Fig~\ref{fig:Fig2}. First, the scanned image is converted to a binary image. From the binary image, the isoelectric lines of the ECG signal are found, and from the position of these lines, the image of each lead is separated. Since the binary image contains noise and artifacts, a noise removal algorithm is required. Finally, from the signal image of each lead, the ECG signal is extracted and concatenated to form a time series ECG signal.

    \paragraph{Image binarization}
    ECG scans are converted from colored to binary by Otsu's method~\cite{yousefi2011image} to facilitate further processing.
    
\begin{figure}[hbt!]
          \centering
          \subfloat[]{
          \includegraphics[width=0.45\linewidth]{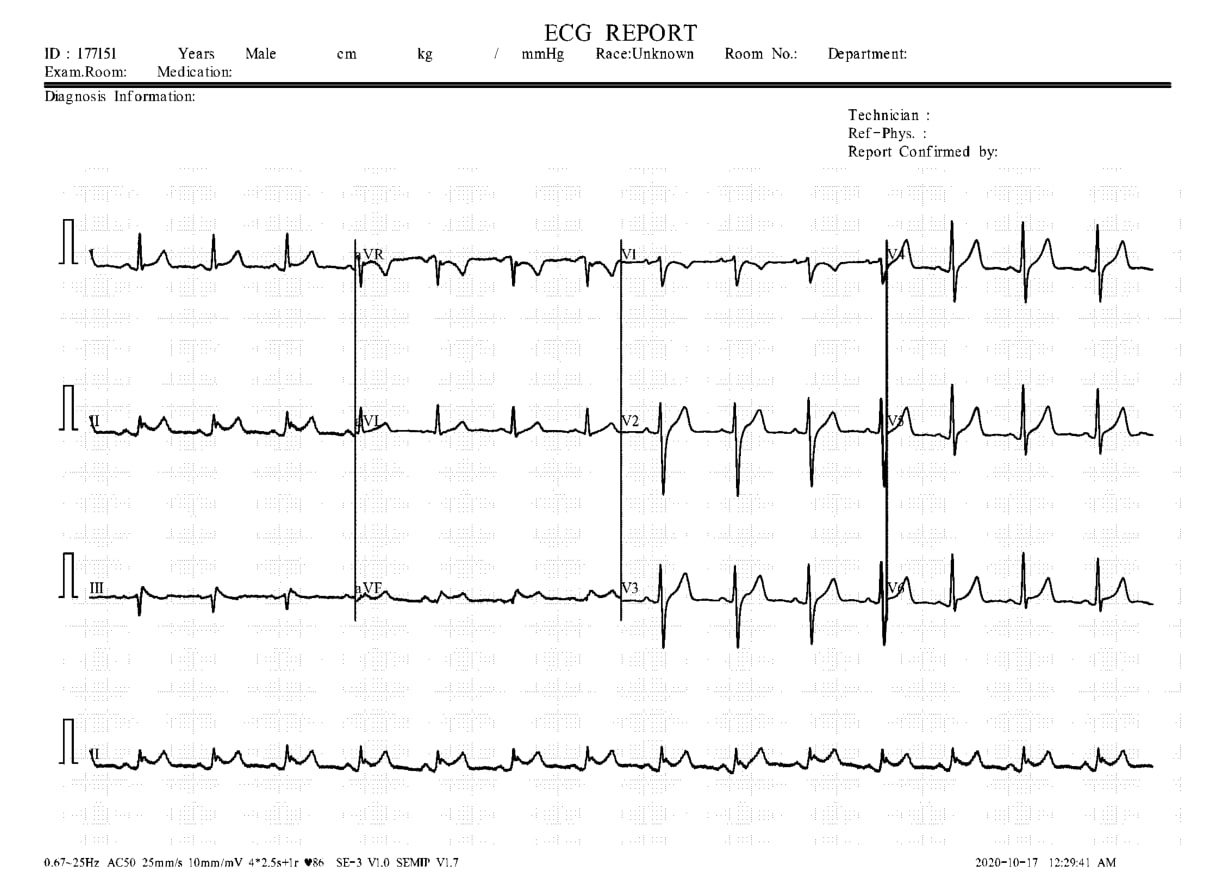}
          \label{subfig: Subfig3a}}
          \subfloat[]{
          \includegraphics[width=0.45\linewidth]{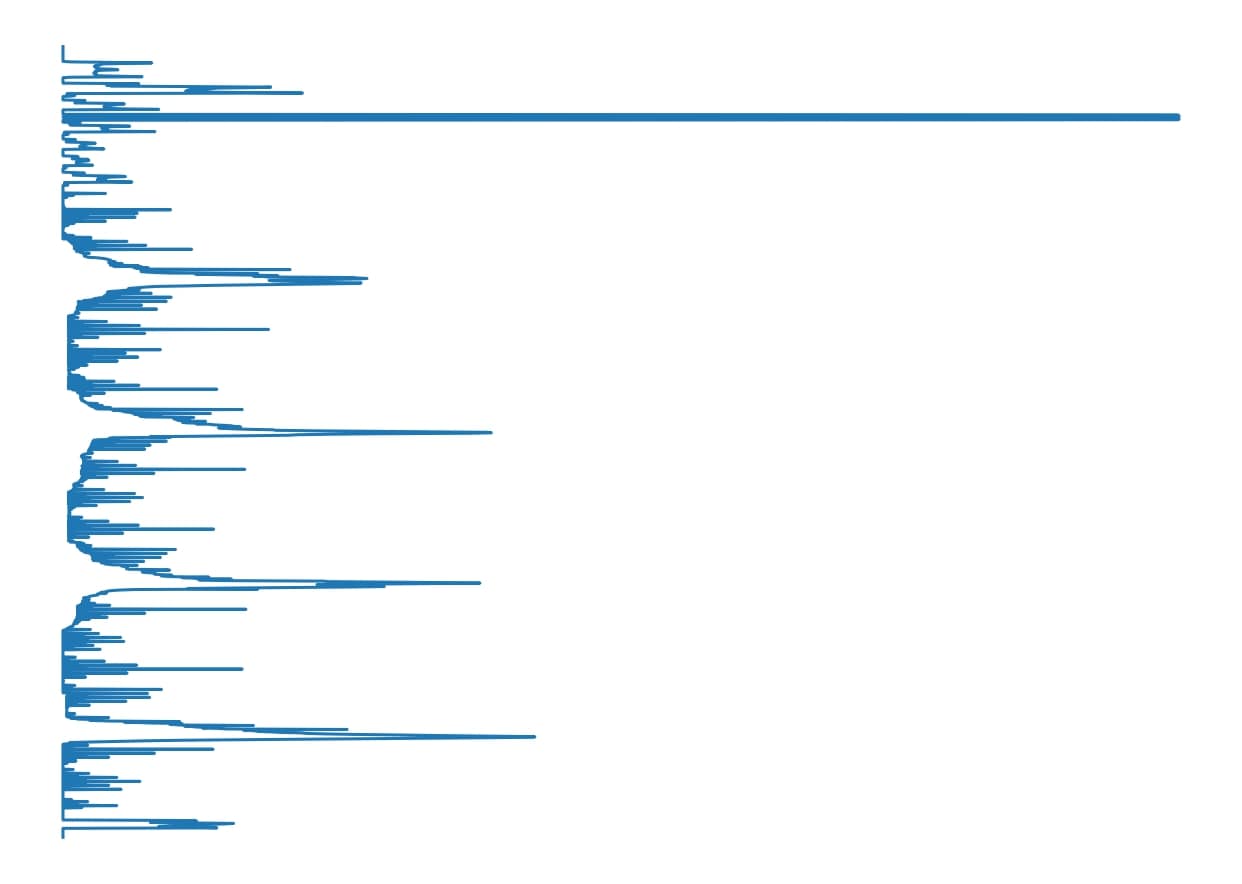}
          \label{subfig: Subfig3b}} \\
          \subfloat[]{
          \includegraphics[width=0.45\linewidth]{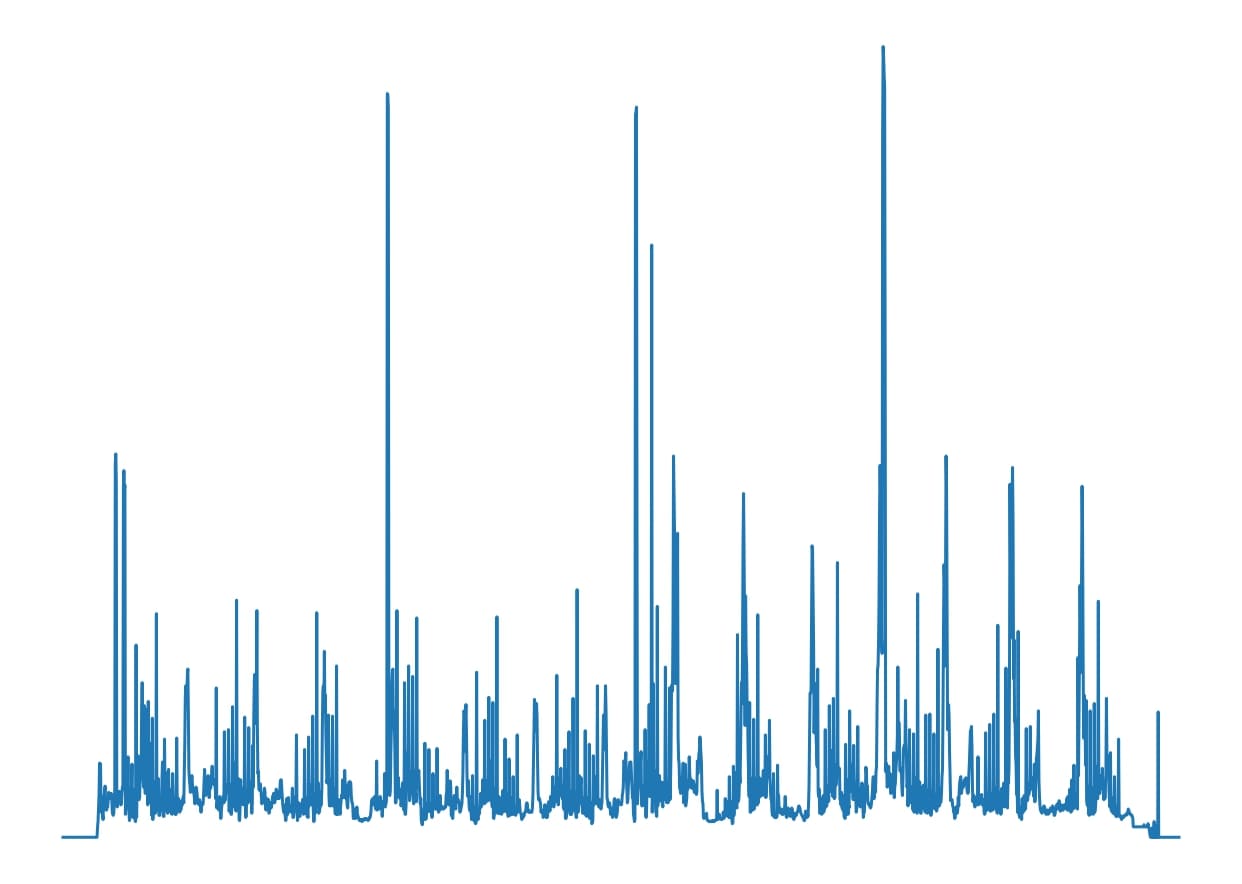}
          \label{subfig: Subfig3c}}
          \subfloat[]{
          \includegraphics[width=0.45\linewidth]{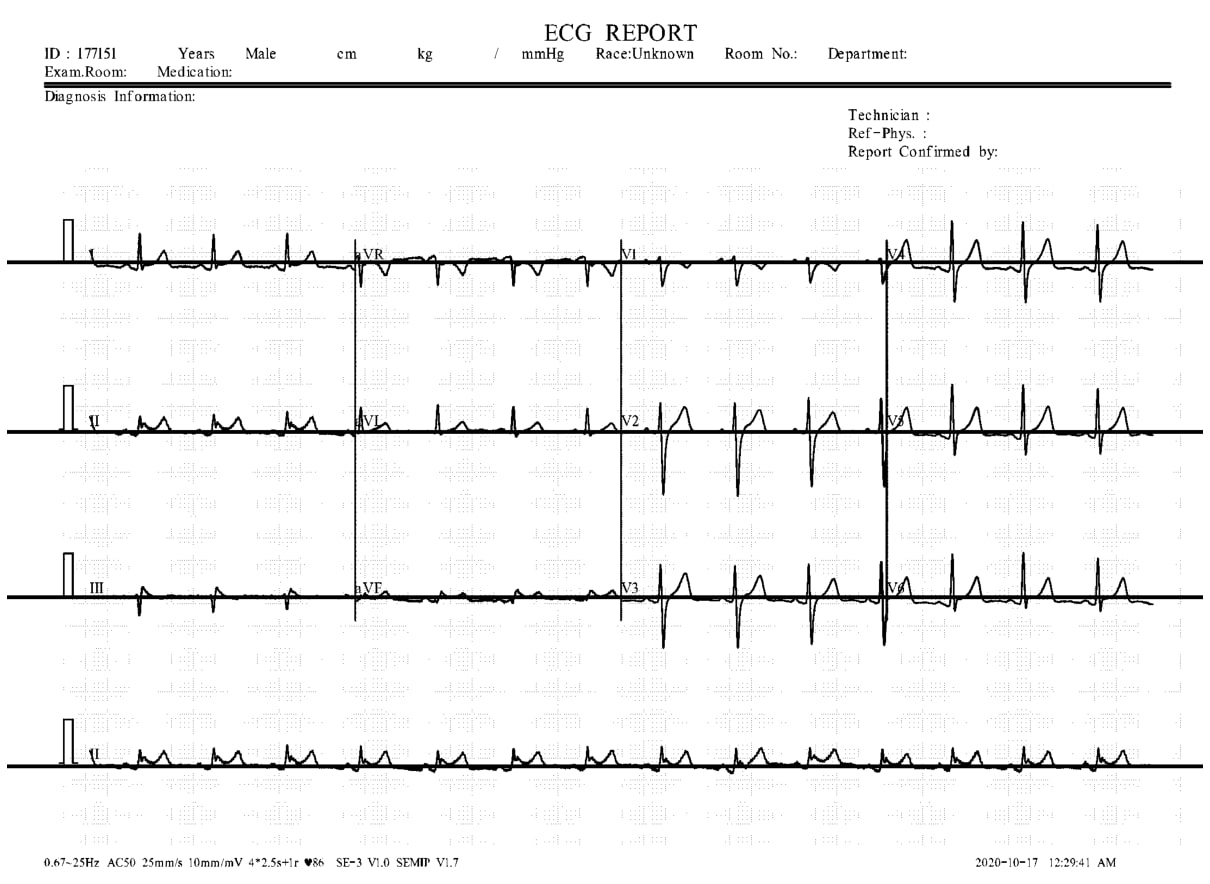}
          \label{subfig: Subfig3d}}
          \caption{{\bf Vertical and horizontal histograms of the ECG image.} {\em (b)} and {\em (c)} are the vertical and horizontal histograms of image {\em (a)}, respectively. {\em (b)} has peaks corresponding to positions of the horizontal lines at the top of the paper and the isoelectric lines of the ECG signal. {\em (c)} has three peaks corresponding to lines separating the leads. In {\em (d)}, the black line depicts the positions of all isoelectric lines.}
          \label{fig:Fig3}
        \end{figure}

    \paragraph{Isoelectric lines detection}
It is critical to find the position of the isoelectric lines in the ECG recording, which can deduce the area of each lead on the image. Due to the characteristic of the isoelectric lines, when projecting the ECG image in the vertical direction, the data points (black points as shown in Fig~\ref{subfig: Subfig3a}) will be concentrated with high intensity at their position. As shown in Fig~\ref{subfig: Subfig3b}, the vertical histogram of the image 
        has peaks at the positions of the isoelectric lines. Thus, finding the position of the isoelectric lines can be done by finding the peaks of the histogram. As shown in Fig~\ref{subfig: Subfig3d}, the isoelectric lines are detected (the black horizontal lines) and shared by four signals corresponding to four ECG leads each.
        
\begin{figure}[hbt!]
      \centering
      \includegraphics[width=\linewidth, scale=0.8]{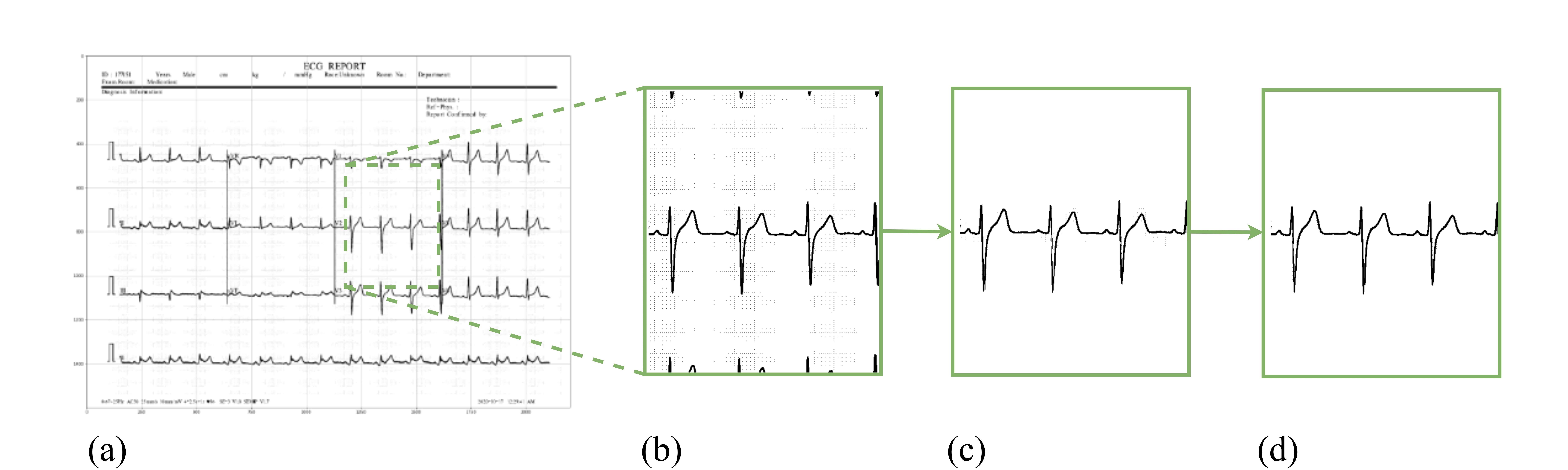}
      \caption{{\bf Image processing for ECG signal extraction.} {\em (a)} is a binary image of the ECG signal, {\em (b)} is an image of the lead V2 signal cropped from {\em (a)}, {\em (c)} is the image of signal after removing noise and artifacts. In {\em (c)}, there are  small noise spots around the signal. These spots were filtered out in {\em (d)}, leaving only the ECG signal image.}
      \label{fig:Fig4}
    \end{figure}

    \paragraph{Lead seperation}
        From the position of the detected isoelectric lines, we separate the image of each lead by zoning vertically from the upper isoelectric line minus L pixels to the lower isoelectric line plus L pixels (L depends on the characteristics of ECG images, specifically the thickness of ECG lines). The boundary between leads sharing an isoelectric line is detected based on the horizontal histogram of the image. Fig~\ref{subfig: Subfig3c} shows the horizontal histogram of the image in Fig~\ref{subfig: Subfig3a}. Due to the presence of characters referring to the name of leads in the ECG paper, the segment of image containing those characters is cropped. 
        Once the vertical and horizontal boundaries of the leads found, we crop the image of each lead for further processing. Fig~\ref{fig:Fig4}{\em a} and {\em b} are instances of an ECG printout and image of lead V2 cropped from it.

    \paragraph{Noise and artifacts removal}
        The binary image is embedded in noise and artifacts from the grid and adjacent signals (as shown in Fig~\ref{fig:Fig4}{\em b}). We built a simple yet effective algorithm that separates the image of the ECG signal from noise and artifacts. The first step involves initializing a point that belongs to the isoelectric line and has a pixel value of 0. This point has coordinate $(n, b)$ ($b$ is the y-intercept of the isoeletric line). Pixels satisfying the following three conditions i) located on the same column as the starting point; ii) having a value of 0 and iii) less than N pixels away from the starting point; are retained. Pixels with a value of 0 in the $n^{th}$ column of the image and are less than N pixels away from the isoelectric line are considered pixels of the ECG signal in that column. Similar to L, the value of N depends on the characteristics of ECG images, specifically the thickness of ECG lines. The maximum coordinate $(c_{max})$ and the minimum coordinate $(c_{min})$ of pixels 0 in that column are calculated. In the next column, pixels with coordinates from $c_{min} - N$ to $c_{max} + N$ are considered; and among them, pixels with value 0 are retained. $c_{max}$ and $c_{min}$ values of this column are again calculated, and the search for ECG pixels in the next column is carried out.  
        
\begin{figure}[hbt!]
          \centering
          \subfloat[]{
          \includegraphics[width=0.33\linewidth, scale=0.2]{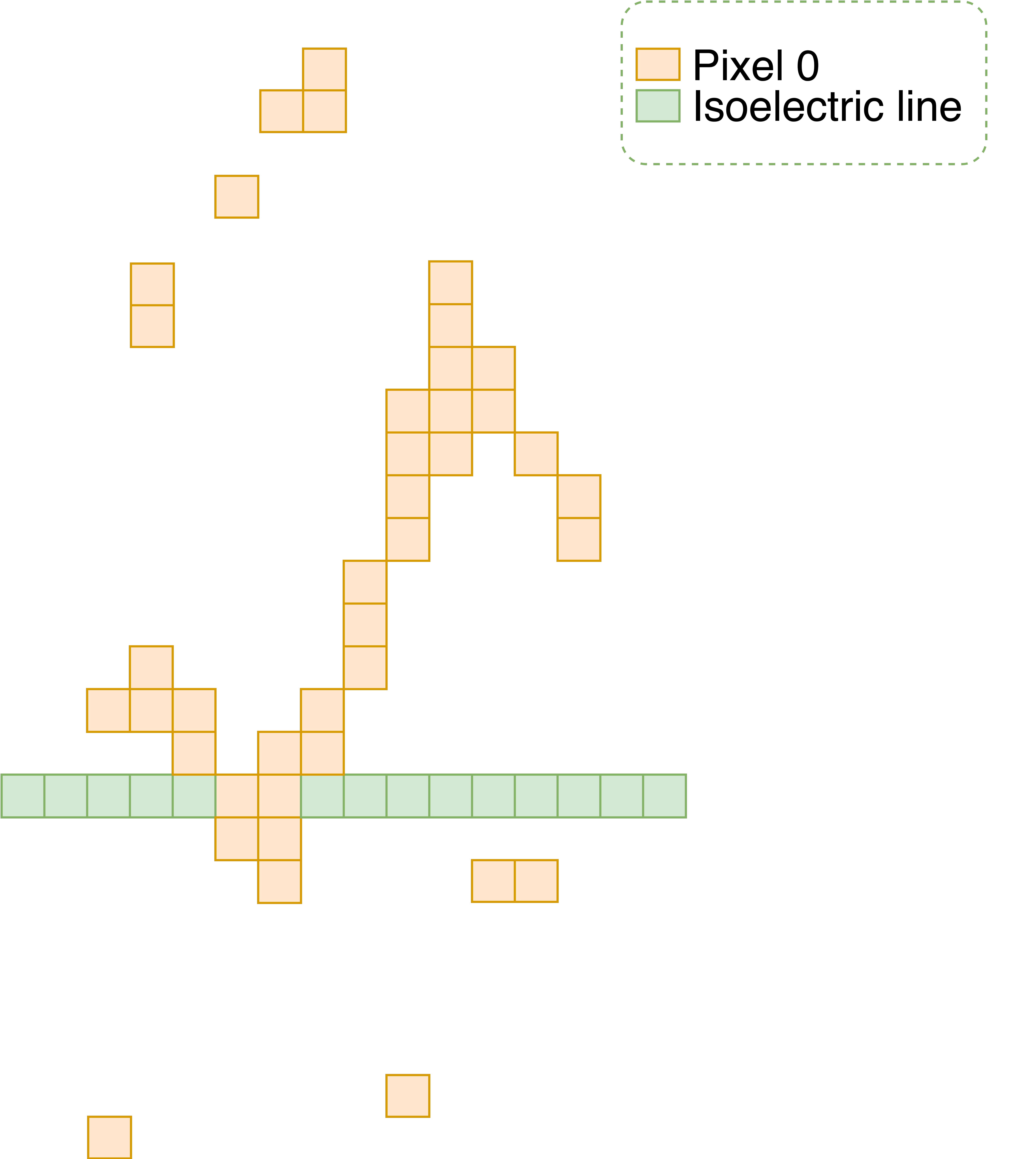}
        }
          \subfloat[]{
          \includegraphics[width=0.33\linewidth, scale=0.2]{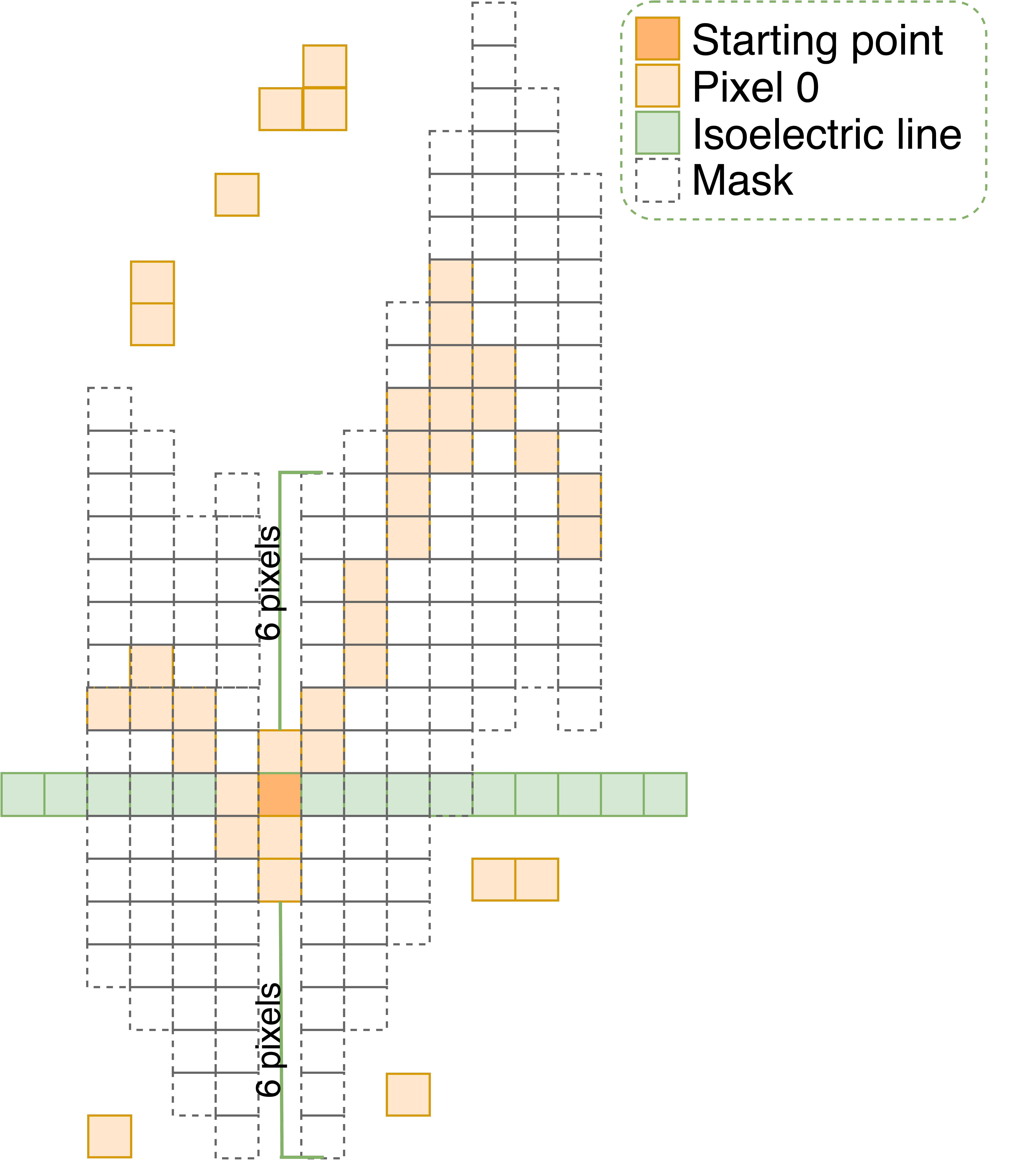}
        }
          \subfloat[]{
          \includegraphics[width=0.33\linewidth, scale=0.2]{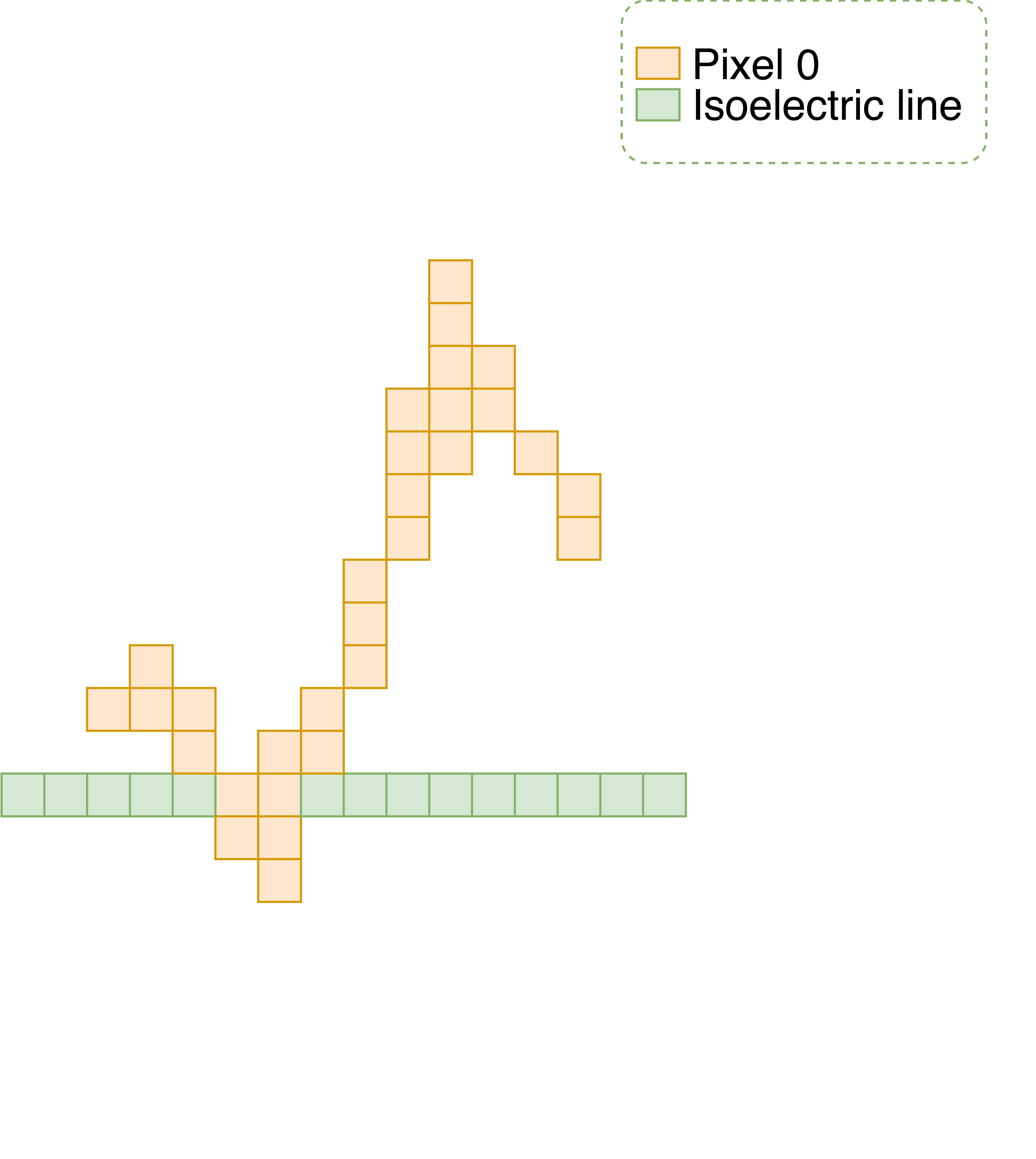}
        }
          \caption{{\bf ECG pixel tracking algorithm} {\em (a)} Signal image with noise and artifacts; {\em (b)} ECG pixel tracking algorithm with $N = 6$; {\em (c)} Image of signal without noise and artifacts.}
          \label{fig:Fig5}
        \end{figure}

        From the starting point, we track in two directions: to the left edge of the image and to the right edge of the image. At the end of the process, only ECG pixels are retained. The noise removal algorithm is detailed in Algorithm~\ref{algorithm1} and Fig~\ref{fig:Fig5}. Fig~\ref{fig:Fig4}{\em c} is  the result of the ECG tracking algorithm applied to Fig~\ref{fig:Fig4}{\em b} with $N=10$. This image contains no noise and artifacts, and is the input for the ECG signal extraction.
        
        
        \begin{algorithm}
        \scriptsize{
            \caption{Noise and artifacts removal algorithm}
            \begin{algorithmic}
                \Ensure Separate the image of ECG signal from noise and artifacts in $img$ and save it to $img\_copy$
                \State $W =$ $img$.shape[1] 
                \State $img\_copy =$ 255*np.ones\_like($img$) 
                \State Initialize $n$ \Comment{Random point for ECG image tracking}
                \State $col = img[b-N:b+N, n]$ \Comment{b: baseline position}
                \State $col =$ np.where($col$)
                \State $img\_copy[col+c_{min}, n] =$ 0
                \State $c_{max} =$ max($col$)
                \State $c_{min} =$ min($col$)
                \For{$i$ in range($n$+1, $W$)}
                    \State $col = img[c_{min}-N : c_{max}+N, n]$
                    \State $col =$ np.where($col$==0)
                    \State $img\_copy[col+c_{min}, n] =$ 0
                    \State $c_{max} =$ max($col$)
                    \State $c_{min} =$ min($col$)
                \EndFor
                \For{$i$ in range($n$-1, -1, -1)}
                    \State $col = img[c_{min}-N : c_{max}+N, n]$
                    \State $col =$ np.where($col$==0)
                    \State $img\_copy[col+c_{min}, n] =$ 0
                    \State $c_{max} =$ max($col$)
                    \State $c_{min} =$ min($col$)
                \EndFor
            \end{algorithmic}
            \label{algorithm1}}
        \end{algorithm}
                
    \paragraph{Signal extraction} The image containing only the ECG pixels is the subject of the image-to-signal conversion.  
    
    Extraction of 1D ECG signals from images is conducted by scanning the image horizontally. The amplitude of the ECG signal at a specific column is equal to the average of the pixel 0 positions of that column minus the isoelectric line position. For columns missing pixel 0, the signal amplitude is taken as the value of the signal point immediately preceding it. Once the scanning is done, the length of the resulting signal is equal to the number of image's columns.
    
\begin{figure}[hbt!]
      \centering
      \subfloat[]{
      \includegraphics[width=0.48\linewidth]{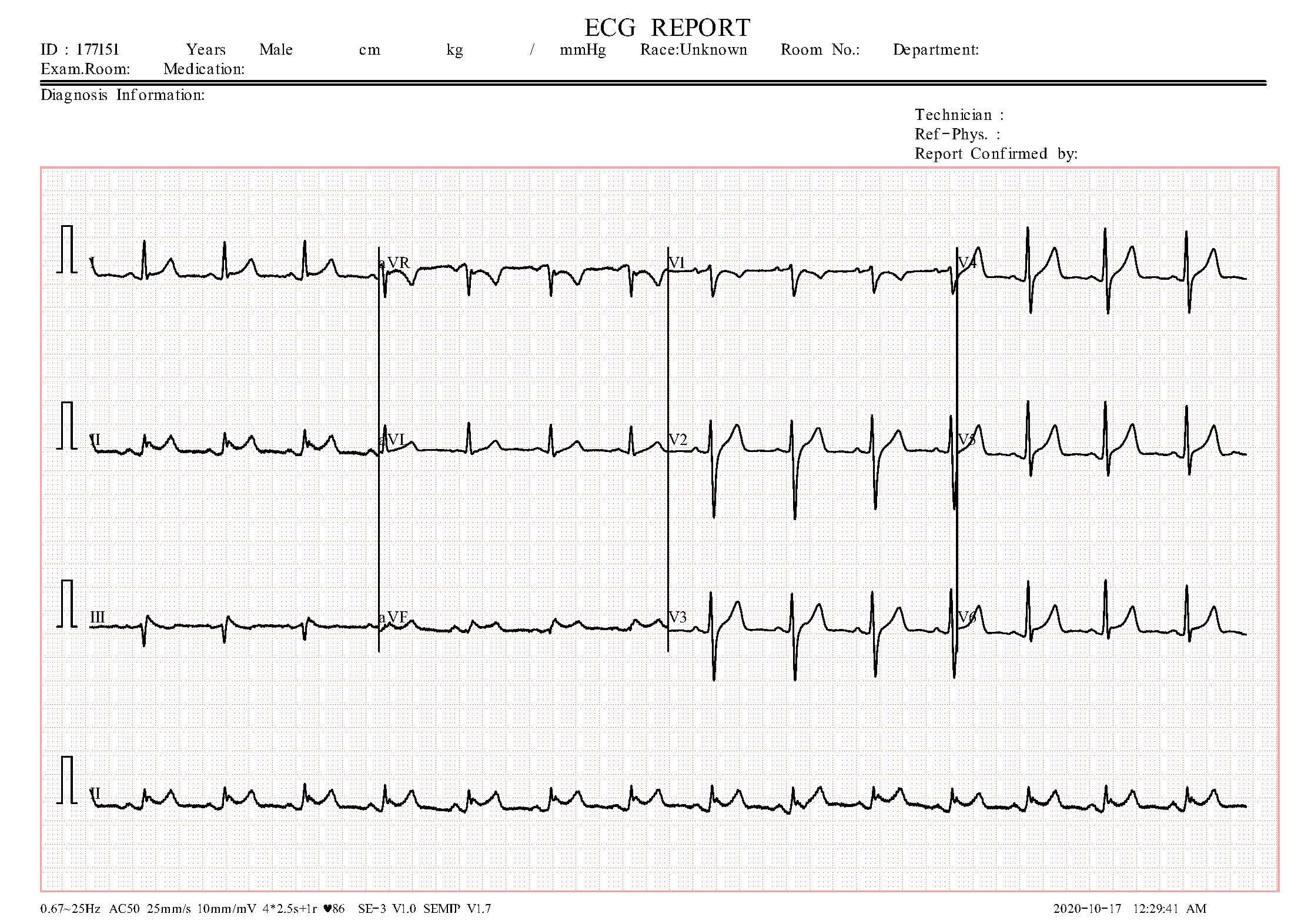}
      \label{subfig:Subfig1a}}
      \subfloat[]{
      \includegraphics[width=0.48\linewidth]{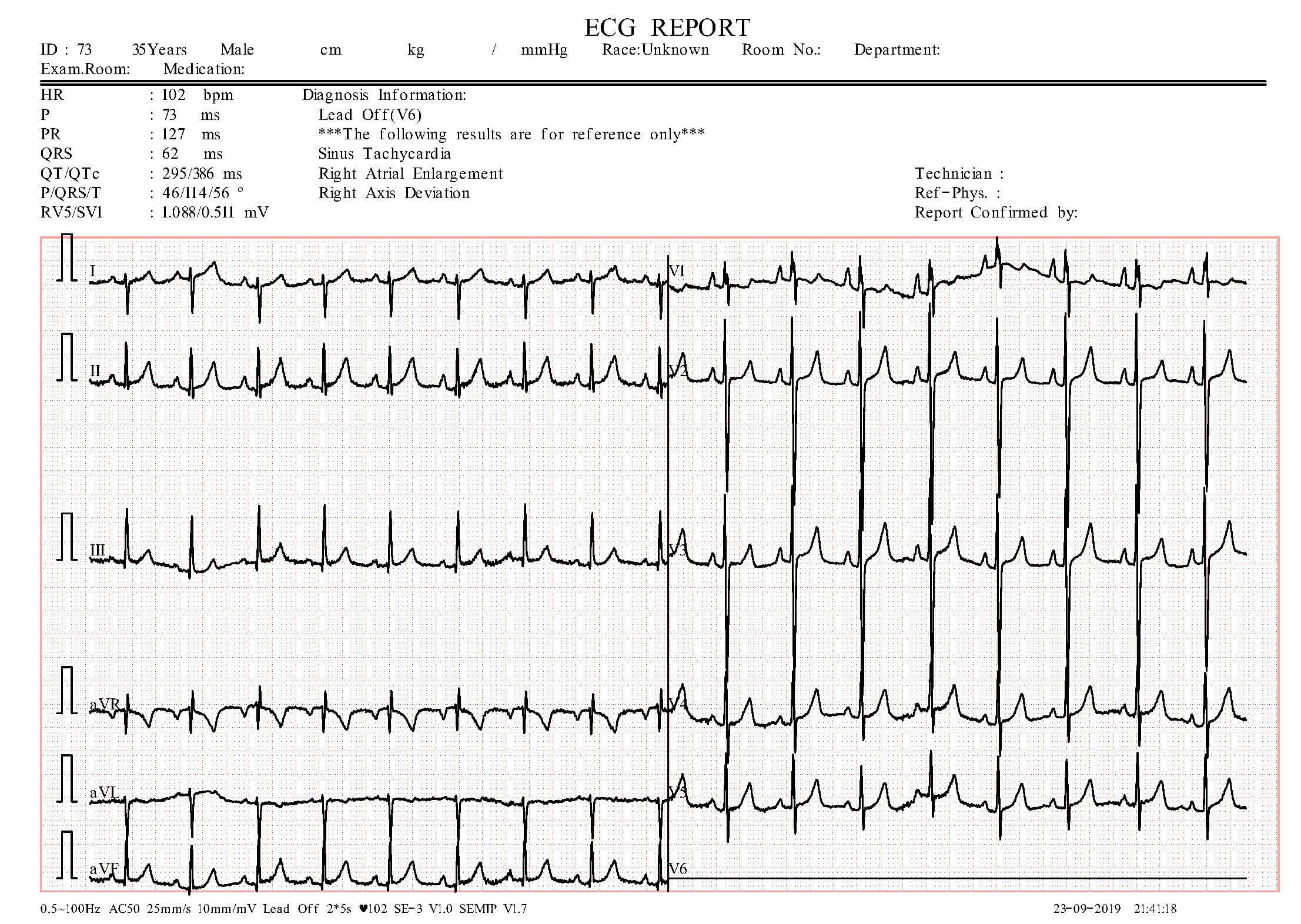}
      \label{subfig:Subfig1b}}
      
      \vspace{-0.4cm}
      \subfloat[]{
      \includegraphics[width=0.48\linewidth]{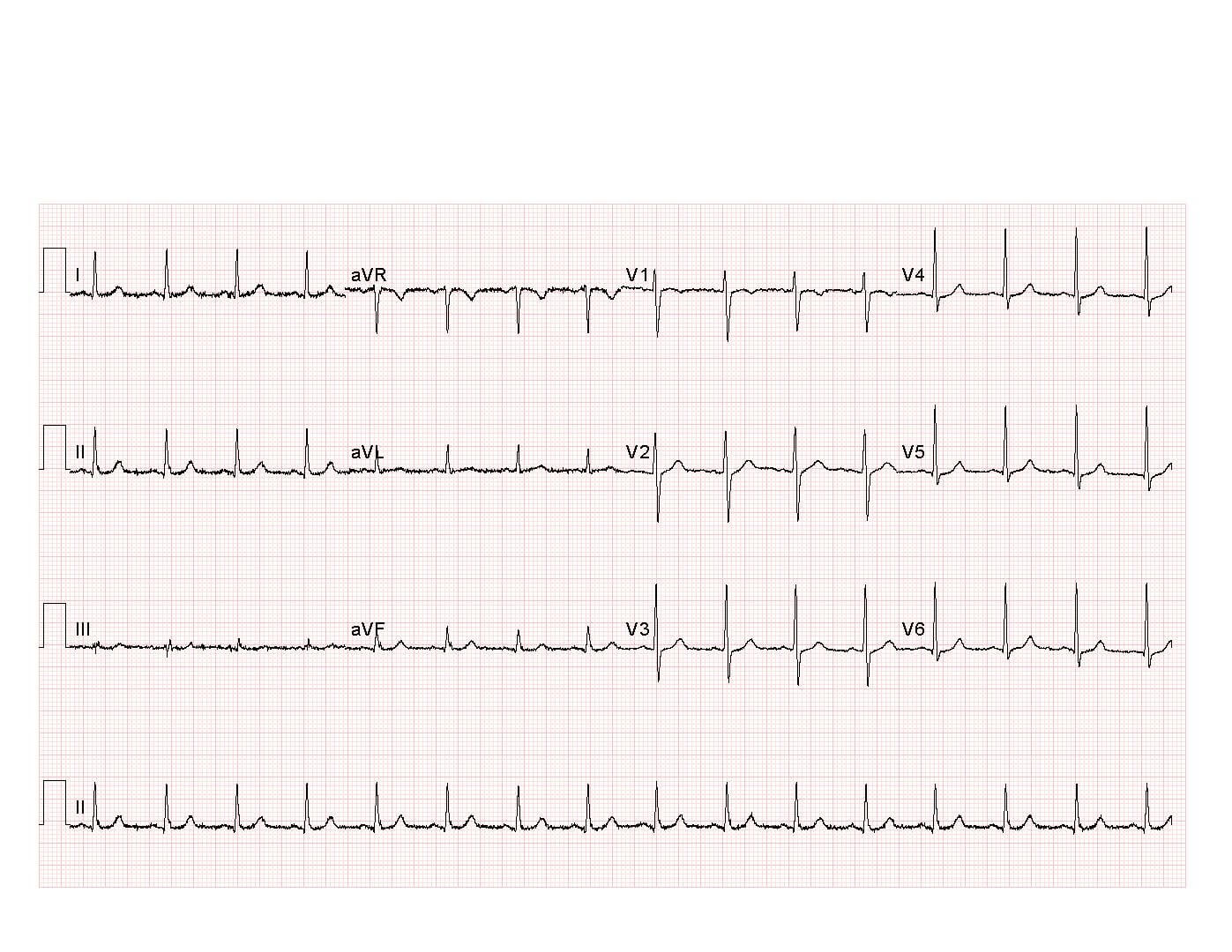}
      \label{subfig:Subfig1c}}
      \subfloat[]{
      \includegraphics[width=0.48\linewidth]{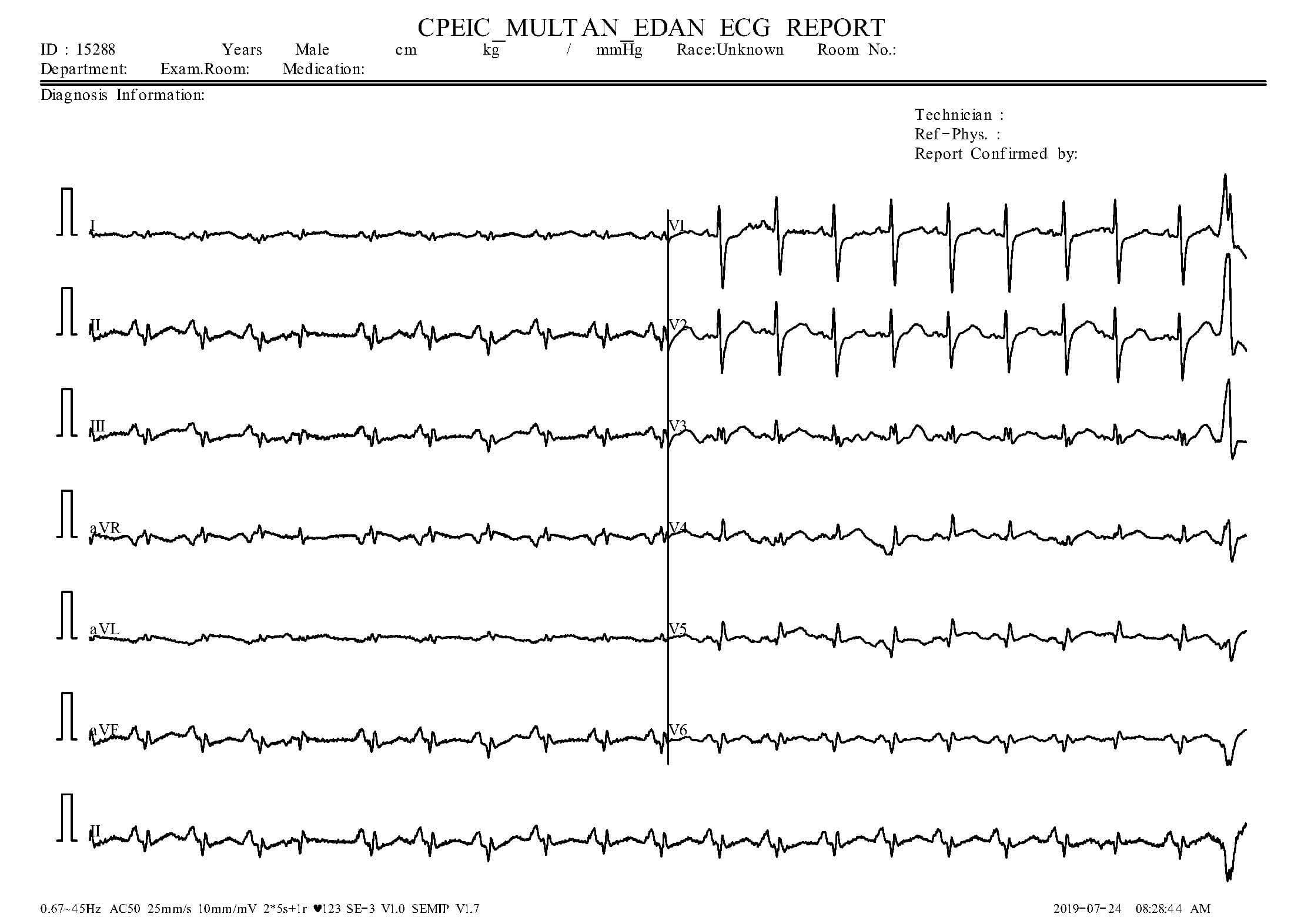}
      \label{subfig:Subfig1d}}
      \caption{{\bf Four types of 12-lead ECG reports in the COVID-19 ECG images dataset}} 
      \label{fig:Fig1}
    \end{figure}

\begin{figure}[hbt!]
      \centering
      \subfloat[]{
      \includegraphics[width=0.45\linewidth]{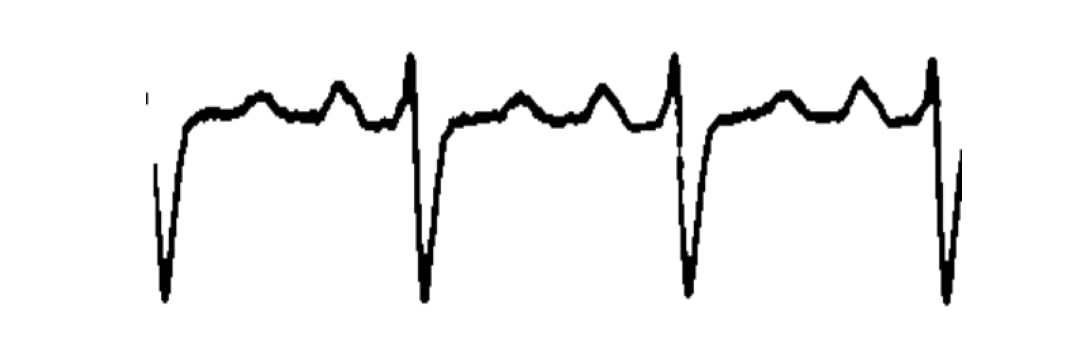}
      \label{subfig:Subfig6a}}
      \subfloat[]{
      \includegraphics[width=0.5\linewidth]{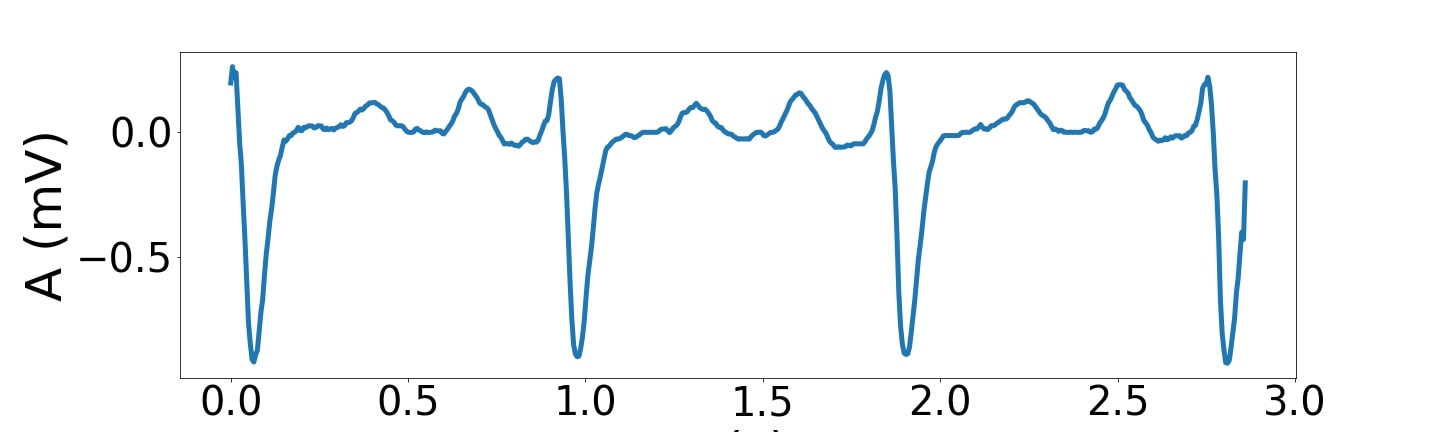}
      \label{subfig:Subfig6b}}
      \caption{\textbf{\textit{(a}) Image of an ECG signal in lead II after preprocessing and \textit{(b)} the resulting signal after image-signal conversion.}}
      \label{fig:Fig6}
    \end{figure}

    Each ECG printout has reference pulses in the form of a square pulse (as shown in Fig~\ref{fig:Fig1}). In Fig~\ref{subfig:Subfig1a}, Fig~\ref{subfig:Subfig1b} and Fig~\ref{subfig:Subfig1d}, the reference square pulses have a width corresponding to 0.1 s and a height corresponding to 0.5 mV. In Fig~\ref{subfig:Subfig1c} the reference pulse is 0.2 s and 0.5 mV for width and height, respectively. Determining the width and height of these square pulses makes the image-to-signal conversion consistent in time and amplitude. In addition, the scale conversion also helps to display the signal accurately in terms of amplitude and time. In this study, we convert all the records to the same sampling frequency of 200 Hz. Fig~\ref{subfig:Subfig6a} depicts an ECG signal image, and Fig~\ref{subfig:Subfig6b} depicts the signal extracted from Fig~\ref{subfig:Subfig6a}.
    
    \paragraph{Signal quality assessment}
    Once the 1D ECG data have been obtained, we assess its quality to guarantee that the image-to-signal conversion algorithm functions properly and that the resulting signal can be used as input for the machine learning model. 
    After randomly selecting 50 samples from each category, we hand-labeled the R peaks of the lead II signal in these images and regarded the resulting positions as ground truths to evaluate our algorithm against.
    The selected samples are also fed into our proposed image-to-signal conversion, and the generated signals were exposed to NeuroKit2's~\cite{makowski2021neurokit2} R peak detection algorithm to obtain the positions of R peaks.
    Finally, we evaluate our algorithm by comparing its output against the ground truths, specifically by quantifying the difference in the RR intervals (the time elapsed between two successive R-waves) using the mean absolute error (MAE) metric.

\section*{Experiments}
\label{section:experiments}
\subsection*{Experimental setups}
\paragraph{Dataset}
    \textit{COVID-19 ECG images dataset} --- a publicly available dataset of ECG images of cardiac and COVID-19 patients was used in this study. Khan \textit{et al.} from the University of Management and Technology shared the dataset to Mendeley Data~\cite{khan2021ecg}. The dataset contains photos of 1937 paper-based ECG reports. ECG records for 250 COVID-19 patients, 77 MI patients, 548 individuals with irregular heartbeats (recovered from COVID-19 or MI), 203 patients with MI history (RMI), and 859 people without any cardiac abnormalities were analyzed by experts. The provided dataset is the first ECG dataset for COVID-19 disease that has been shared and is completely anonymized. We did not seek approval from an ethics committee and participants' consent for this study because (i) the previous study during which the dataset was collected already obtained necessary approval, (ii) the dataset is completely anonymized and contains no personal information which can be used to identify patients, and (iii) no further data are collected during this study.
    
    The ECG images in the dataset are from a 12-lead system (I, II, III, aVR, aVL, aVF, V1, V2, V3, V4, V5, and V6) at a sampling rate of 500 Hz. ECG data were gathered using an EDAN SE-3 series 3-channel electrocardiograph, and some of the signals were applied with a 0.67–25 Hz bandpass filter, while others were processed with a 0.5–100 Hz bandpass filter and a 50 Hz notch filter, according to the paper-based ECG reports.  
    
    The dataset has some limitations. In particular, the dataset is considerably imbalanced, with 859 samples of normal ECG (the class with the most samples) and 77 samples of MI ECG (the class with the fewest samples). Besides, the dataset consists of four different types of reports, which are different in (i) image size and resolution; (ii) ECG report template; (iii) grid background, and (iv) reference pulse. Four types of ECG printout templates are shown in Fig~\ref{fig:Fig1}. Using the signal extracted from the ECG image is not only reduces the computational cost but also removes disparities in ECG printout templates, making the prediction more objective. 
\paragraph{Experiments}
    To evaluate the capability of diagnosing COVID-19 based on ECG data, three classification tasks are covered:
    (i) COVID-19 and normal ECG;
    (ii) COVID-19 and others;
    (iii) COVID-19, normal and others. As the number of samples in MI and RMI classes is relatively small, the merging of MI, RMI, and abnormal into a class could alleviate the effects of data imbalance. The number of images used in each problem is detailed in Table~\ref{tab:ours}.

    To evaluate the effectiveness of this method, we also compare its performance against the results of previous studies performed on the \textit{COVID-19 ECG images dataset}. We simulate the data splitting of previous studies and execute similar classification tasks~\cite{ozdemir2021classification,irmak2022covid,attallah2022ecg}.
    
    \paragraph{Implementation details}
    In this study, all experiments were conducted using Intel core i9-10900X CPU @ 3.70GHz, RAM 31 GB hardware and NVIDIA GeForce RTX 3080 Ti GPU. We trained 110 epochs for each model, with a mini-batch size of 16, and evaluated model performance after each epoch. Checkpoint with the highest F1-score against the test set was considered the best model for each training procedure. As the backbone of the COVID-19 vs. Normal classifier, we apply multiple architectures, then select the model with the best performance to train other classification tasks. The utilized architectures are SEResNet18~\cite{hu2018squeeze}, ResNet18~\cite{wang2017time}, ResCNN~\cite{zou2019integration}, Sequence Stroke Net~\cite{liu2020stroke}, LSTM Fully Convolutional Network (LSTM FCN)~\cite{karim2017lstm}, GRU Fully Convolutional Network (GRU FCN)~\cite{grufcn}, InceptionTime~\cite{inception}, XceptionTime~\cite{xception} and Time Series Transformer (TST)~\cite{zerveas2021transformer}. 
    We adopted Adam optimizer~\cite{kingma2014adam} ($\beta 1 = 0.9$, $\beta 2 = 0.999$ and learning rate = 1e-3), cooperating with Cosine annealing learning rate~\cite{loshchilov2016sgdr} with the maximum number of iteration set to 25 and the learning rate stop decaying at $100^{th}$ epoch. As a criterion for training, the label smoothing cross-entropy loss function with a smoothing factor of 0.1 was used. We selected the most suitable hyper-parameters using simple grid search strategy and experimental results.
    
    To evaluate the performance of the models, a nested cross-validation method was employed for each trial. The dataset was first divided into 5 folds, with 1 set used for testing and the rest used for training and validation. In each iteration, we used 4-fold cross-validation to further split the data between training and validation. Classification performance was evaluated for each fold, and then the average classification performance of each model was calculated. To ensure that the data distributions in training, validation and test sets were the same as in the original dataset, stratified splitting was adopted for both cross-validation steps.
    
    We used all of the samples in the dataset that are relevant to the considered classes for each classification task. The number of images used for each task is listed in Table~\ref{tab:ours}

\subsection*{Evaluation metrics}
    To evaluate the quality of the ECG signals extracted from the images, we used mean absolute error (MAE) as the evaluation metric. The mean of the RR intervals in the ground truth and the resulting signal is calculated, then MAE value is calculated. Equation~\ref{eq:mae} gives the formula for computing the MAE.
    \begin{equation}
        \label{eq:mae}
        MAE = \frac{\sum_{i=1}^{N}|mRR_i - \hat{mRR_i}|}{N}
    \end{equation}
    Where $N$ is total number of ECG images; $mRR_i$ is the mean of RR intervals calculated from $i^{th}$ image; $\hat{mRR_i}$ is the mean of RR intervals of the corresponding signal.
    
    In terms of the CNN model, model performance is assessed utilizing well-known evaluation metrics such as Accuracy, F1-score, Specificity, Sensitivity, and AUC (Area Under the Curve of the Receiver Operating Characteristic). 
    
    
    
    

\section*{Results}
\label{section:results}
\subsection*{Experimental results and implications}
    \paragraph{ECG signal quality assessment}
        Extracted signal quality assessment was evaluated on a certain number of samples in the dataset. The number of samples in each class, as well as the MAE of the RR interval, are listed in Table~\ref{tab:QA}. Despite starting with 50 samples in each class \ref{section:methods}, the RR interval cannot be calculated for samples with only one R peak and thus the reported sample counts were less than 50.
        
        The MAE values range from 11.09 ms to 63.75 ms, as shown in Table~\ref{tab:QA}. Normal class has the lowest MAE (11.09 ms) due to the regular and stable ECG waveform. Class MI has the highest MAE of 63.75 ms, due to the unstable of ECG form of patients with MI disease. The MAE values for COVID-19, Abnormal, and RMI classes are 16.07 ms, 25.73 ms, and 27.57 ms, respectively. Overall, the mean MAE of the tested samples is 28.11 ms, which is 3.7\% of the RR interval for a person with an 80 bpm heart rate (equivalent to a mean RR interval of 750 ms).
            \begin{table}[!ht]
            \centering
            \scriptsize{\caption{\textbf{Evaluation results of ECG signal quality obtained from image-to-signal conversion algorithm.} Evaluation was performed on 226 random samples belonging to 5 classes, with mean absolute error of RR interval of 28.11 ms.}
            \label{tab:QA}
            \begin{tabularx}{0.7\linewidth}{>{\centering\arraybackslash}X*{2}{>{\centering\arraybackslash}X}}
            \hline 
            & \textbf{Number of images} & \textbf{MAE (ms)} \\
            \hline 
            COVID-19 & 42 & 16.07 \\
            MI & 41 & 63.75 \\
            RMI & 48 & 27.57 \\
            Abnormal & 47 & 25.73 \\
            Normal & 48 & 11.09 \\
            \textbf{Total} & \textbf{226} & \textbf{28.11} \\
            \hline
            \end{tabularx}}
        \end{table}
        
    \paragraph{ECG signal classification}
    Table~\ref{tab:model_comparason} compares the COVID-19 vs. normal classification results between different models. With an f1-score of 0.9842 and an AUC of 0.9735, the SEResNet18 model outperforms the others. The XceptionTime model has the best classification accuracy (98.58\%), while the ResNet18 model has the fastest inference time (18.34 ms per sample). 
    According to the obtained results, we choose the SEResNet18 with some modifications as the backbone for other classification tasks. Convolution (Conv) layers are modified with a larger kernel size $(kernel\_size = 7)$ and a stride parameter of 1 in order to capture longer patterns in ECG data. For ECG data in specific and time-series data in general, this technique has been recommended to be more effective~\cite{cheng2021ecg, ismail2020inceptiontime}. 
    
    Table~\ref{tab:ours} covers all of the experimental results for each classification task. Patients positive for COVID-19 were distinguished from normal patients and patients negative for COVID-19 with high accuracy of 98.42\% and 98.50\%, respectively. The classification accuracy and AUC were 83.17\% and 0.8748 in the case of classifying patients with COVID-19, Normal and remaining classes. 
    The majority of misclassified samples include COVID-19 ECGs with identical characteristics to those of normal ECGs, indicating that the ECG waveform of the patients has not changed.
    
\begin{figure}[hbt!]
      \centering
      \subfloat[]{
      \includegraphics[width=0.48\linewidth]{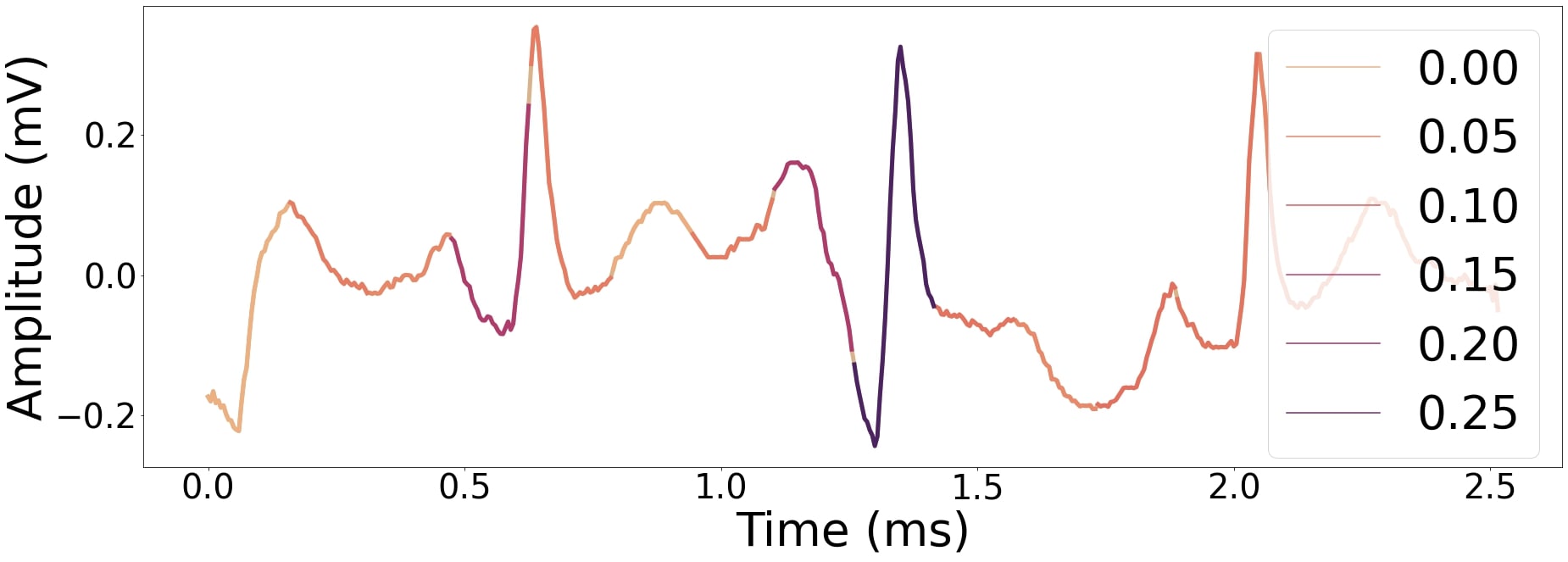}
      \label{subfig:Subfig7c}}
      \subfloat[]{
      \includegraphics[width=0.48\linewidth]{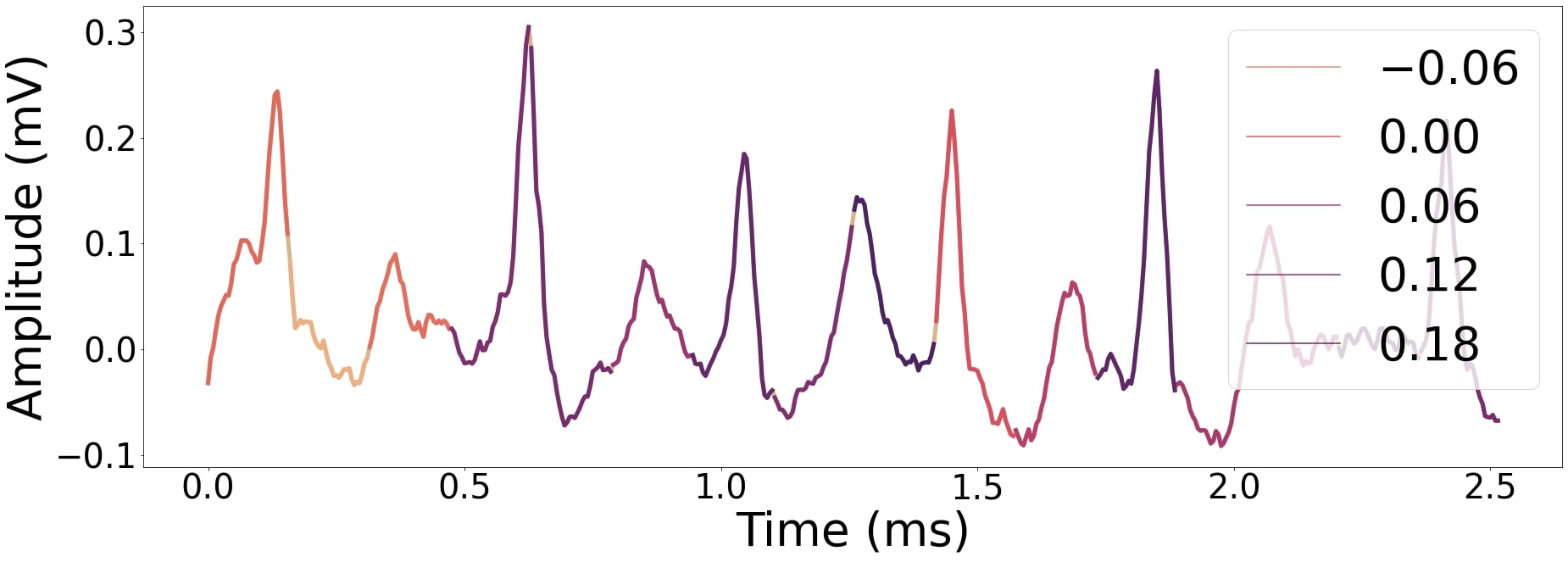}
      \label{subfig:Subfig7d}}
      \vspace{-0.2cm}
      \subfloat[]{
      \includegraphics[width=0.48\linewidth]{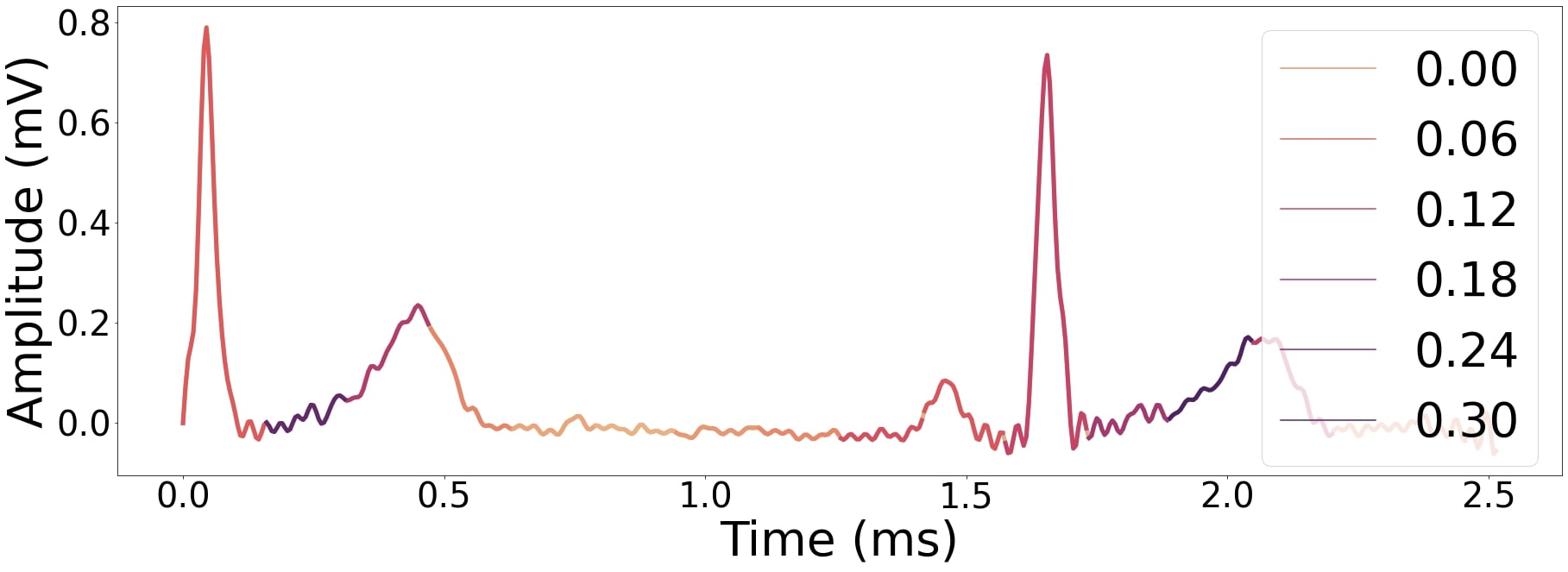}
      \label{subfig:Subfig7e}}
      \subfloat[]{
      \includegraphics[width=0.48\linewidth]{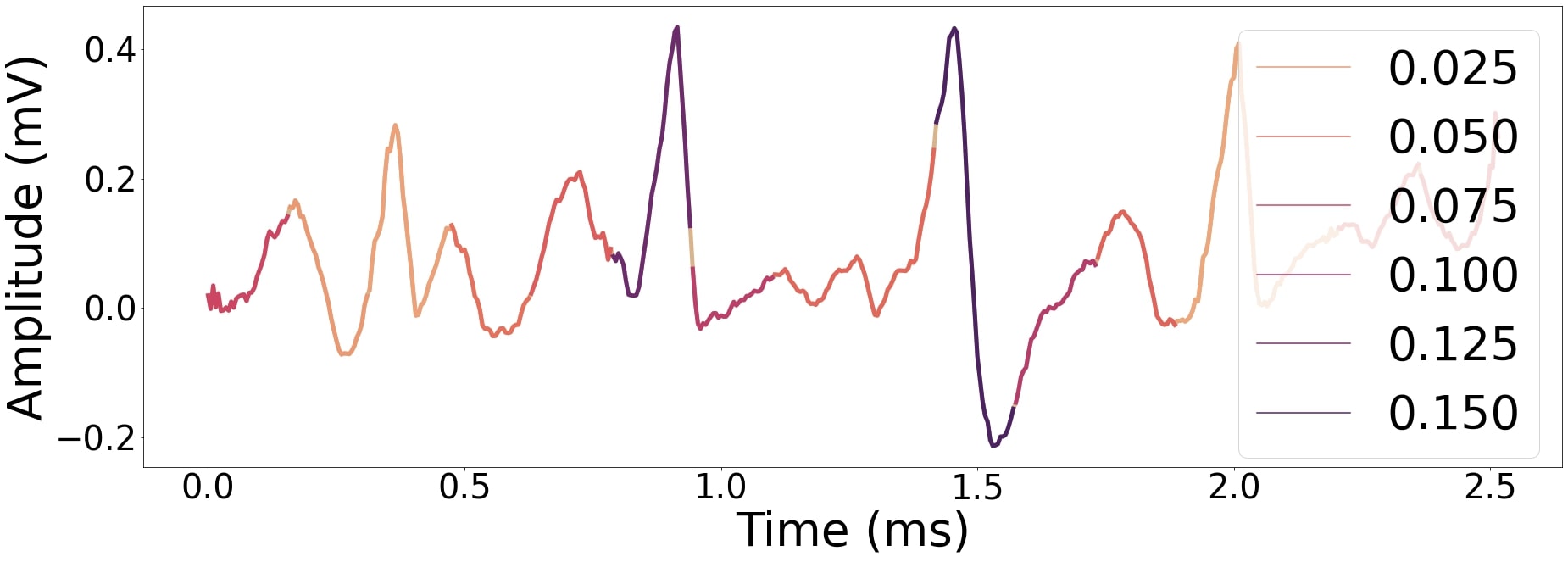}
      
      \label{subfig:Subfig7f}}
      \caption{\textbf{\textit{(a, b)} Grad-CAM visualization of ECG image with COVID-19. \textit{(c, d)} Grad-CAM visualization of ECG image with MI.} Signal segments significant to the classifier are depicted in darker color.}
      \label{fig:Fig7}
    \end{figure}

\begin{figure}[hbt!]
      \centering
      \subfloat[]{
      \includegraphics[width=0.46\linewidth]{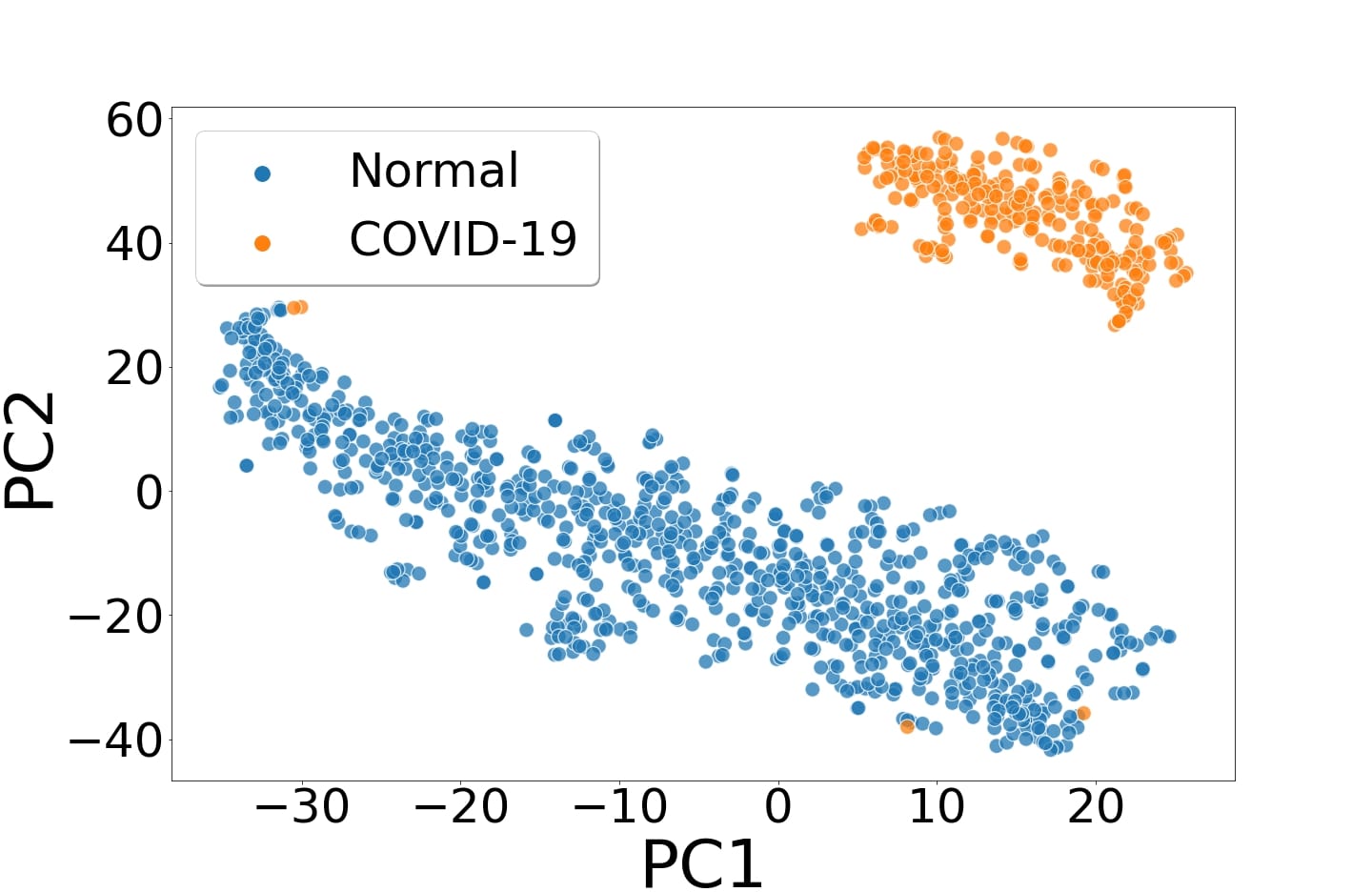}
    }
      \subfloat[]{
      \includegraphics[width=0.46\linewidth]{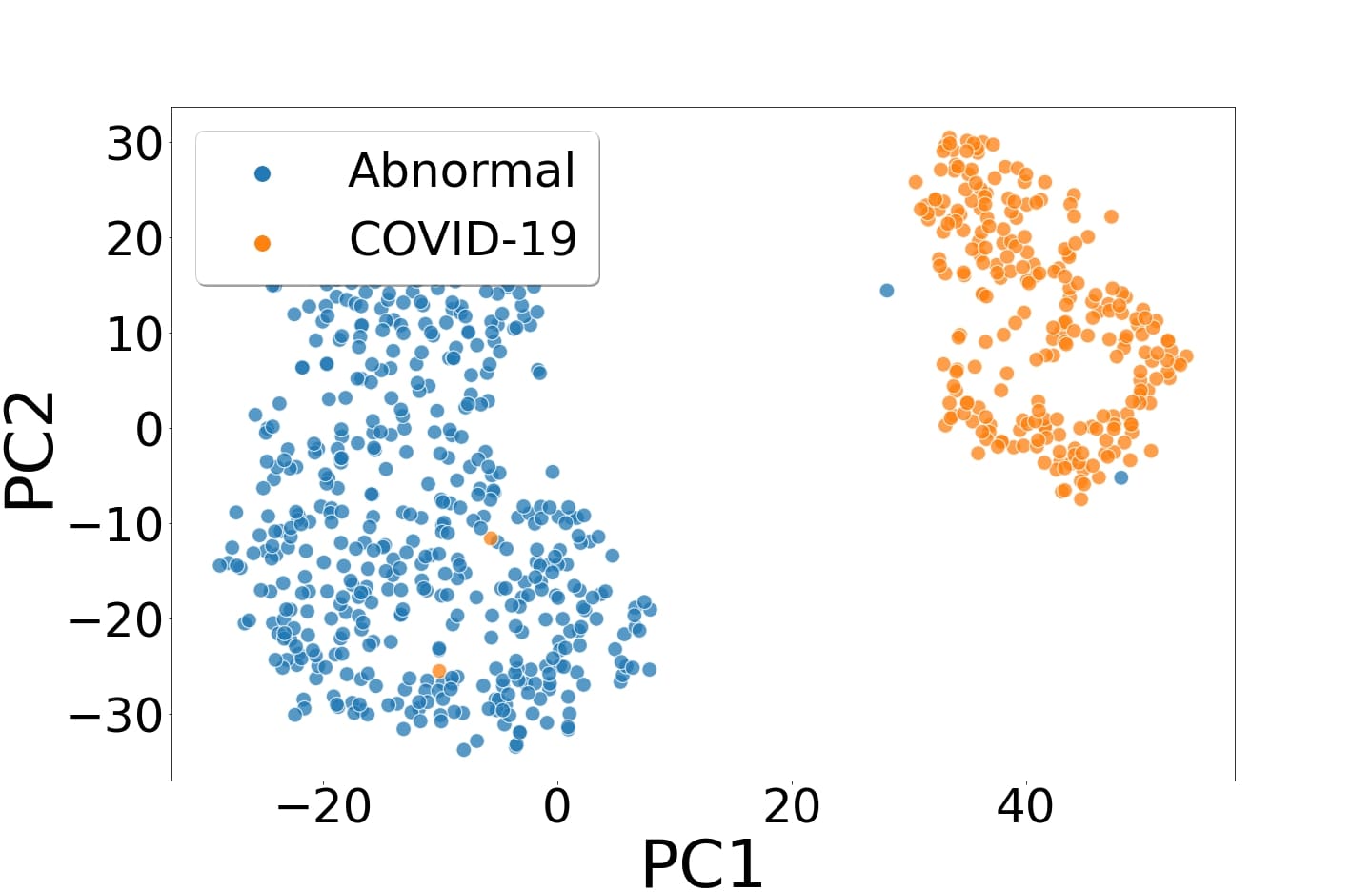}
    }

      \caption{\textbf{t-SNE plots of ECG data of different classes separated at the models' last convolutional layer.}} 
      \label{figure:Fig8}
     \end{figure}

    It is also crucial to discover whether the network is making the decision based on the relevant segment of the ECG signal or somewhere else. For different classes of ECG data, heat maps based on the Grad-CAM technique~\cite{selvaraju2017grad} were constructed. Fig~\ref{fig:Fig7} shows heat maps generated with the best-performing models and Fig~\ref{figure:Fig8} depicts t-SNE plots~\cite{van2008visualizing} to visually highlight the capacity to distinguish between COVID-19 and other conditions.
    
    \begin{table}[!ht]
        \centering
        {\scriptsize
        \caption{\textbf{Classification results     of COVID-19 vs. normal of the different models} based on evaluation metrics are F1-score, accuracy, sensitivity, specificity, AUC and inference time.}
        \label{tab:model_comparason}
        \begin{tabular}{p{0.2\textwidth} p{0.08\textwidth} p{0.08\textwidth} p{0.08\textwidth} p{0.08\textwidth} p{0.08\textwidth} p{0.1\textwidth}}
        \hline 
          &\textbf{F1 score} & \textbf{Accuracy (\%)} & \textbf{Sensitivity (\%)} & \textbf{Specificity (\%)} & \textbf{AUC} & \textbf{Inference time (ms)} \\
        \hline 
         \textbf{SEResNet18} & \textbf{0.9842} & 97.72 & \textbf{97.35} & 98.14 & \textbf{0.9735} & 36.95 \\
        \hline 
         ResNet18 & 0.9738 & 98.20 & 96.89 & 97.94 & 0.9689 & \textbf{18.34} \\
        \hline 
         ResCNN & 0.9761 & 98.35 & 97.09 & 98.18 & 0.9709 & 19.61 \\
        \hline 
         Sequence StrokeNet & 0.9684 & 97.84 & 96.12 & 97.68 & 0.9612 & 24.63 \\
        \hline 
         LSTM FCN & 0.9723 & 98.08 & 96.81 & 97.73 & 0.9681 & 48.80 \\
        \hline 
         GRU FCN & 0.9703 & 97.93 & 96.89 & 97.23 & 0.9689 & 81.23 \\
        \hline 
         Inception Time & 0.9780 & 98.49 & 97.18 & 98.52 & 0.9718 & 81.28 \\
        \hline 
         Xception Time & 0.9792 & \textbf{98.58} & 97.28 & \textbf{98.67} & 0.9728 & 44.40 \\
        \hline 
         TST & 0.9733 & 98.17 & 96.62 & 98.14 & 0.9662 & 32.93 \\
         \hline
        \end{tabular}}
    \end{table}
    
    
    \begin{table}[!ht]
        \centering
        {\scriptsize\caption{\textbf{Results of classifying COVID-19 patients with other groups.} Three classification tasks were completed in order to determine whether ECG signals might be used to diagnose COVID-19.}
        \label{tab:ours}
        
        \begin{tabular}{p{0.02\textwidth}p{0.12\textwidth}p{0.08\textwidth}p{0.08\textwidth}p{0.08\textwidth}p{0.08\textwidth}p{0.08\textwidth}p{0.08\textwidth}p{0.08\textwidth}}
        \hline 
          & \textbf{Classification task} & \textbf{Number of images} & \textbf{Accuracy (\%)} & \textbf{F1-score (\%)} & \textbf{Sensitivity (\%)} & \textbf{Specificity (\%)} & \textbf{AUC (\%)} & \textbf{Inference time (ms)} \\
        \hline 
         \multirow{2}{*}{1} & COVID-19 & 250 & \multirow{2}{*}{98.42} & \multirow{2}{*}{97.72} & \multirow{2}{*}{97.35} & \multirow{2}{*}{98.14} & \multirow{2}{*}{97.35} & \multirow{2}{*}{36.95} \\
          & Normal & 859 &   &   &   &   &   &   \\
        \hline 
         \multirow{2}{*}{2} & COVID-19 & 250 & \multirow{2}{*}{98.50} & \multirow{2}{*}{96.55} & \multirow{2}{*}{95.39} & \multirow{2}{*}{97.93} & \multirow{2}{*}{95.39} & \multirow{2}{*}{66.20} \\
          & Others & 1687 &   &   &   &   &   &   \\
        \hline 
         \multirow{3}{*}{3} & COVID-19 & 250 & \multirow{3}{*}{83.17} & \multirow{3}{*}{85.38} & \multirow{3}{*}{84.81} & \multirow{3}{*}{86.28} & \multirow{3}{*}{87.48} & \multirow{3}{*}{77.56} \\
          & Normal & 859 &   &   &   &   &   &   \\
          & Others & 828 &   &   &   &   &   &   \\
         \hline
        \end{tabular}}
    \end{table}
    
\subsection*{Comparison with state-of-the-arts}
 Table~\ref{tab:compare} summarizes the experimental results and comparisons against previous studies using the same ECG dataset. The proposed method has an accuracy of 0.6\% higher than that of previous studies in the COVID-19 vs. Normal classification problem and up to 6.72\% and 6.43\% higher in the other classification problems in terms of to accuracy and F1-score, respectively. 

\begin{table}[!ht]
    \centering
    \scriptsize{\caption{\textbf{Comparison to state-off-the-arts.} To evaluate the efficacy of our method in comparison to others, data was split in a way that simulates related studies. Dashes (-) represent values not reported in the original paper. Our results are in bold.}
    \label{tab:compare}
    \begin{tabular}{p{0.02\textwidth}p{0.4\textwidth}p{0.16\textwidth}p{0.09\textwidth}p{0.08\textwidth}}
    \hline 
    \textbf{Ref} & \textbf{Classification problem} & \textbf{Method} & \textbf{Accuracy} & \textbf{F1-score} \\
    \hline 
    
     \multirow{4}{*}{\cite{ozdemir2021classification}} & \multirow{2}{*}{\makecell[l]{2-class classification\\(COVID-19 and normal)}} & Hexaxial feature mapping & 96.25\% & 0.9630 \\
    \cline{3-5} 
       &   & \textbf{Ours} & \textbf{96.85\%} & \textbf{0.9685} \\
    \cline{2-5} 
       & \multirow{2}{*}{\makecell[l]{2-class classification\\(COVID-19 and others)}} & Hexaxial feature mapping & 93.30\% & 0.9320 \\
    \cline{3-5} 
       &   & \textbf{Ours} & \textbf{96.78\%} & \textbf{0.9570} \\
    \hline 
    \multirow{4}{*}{\cite{irmak2022covid}} & \multirow{2}{*}{\makecell[l]{2-class classification\\(COVID-19 and normal)}} & Novel CNN & 98.57\% & - \\
    \cline{3-5} 
       &   & \textbf{Ours} & \textbf{99.12\%} & \textbf{0.9870} \\
    \cline{2-5} 
       & \multirow{2}{*}{\makecell[l]{2-class classification\\(COVID-19 and abnormal)}} & Novel CNN & 93.20\% & - \\
    \cline{3-5} 
       &   & \textbf{Ours} & \textbf{95.01\%} & \textbf{0.9398} \\
    
    \hline 
    
     \multirow{4}{*}{\cite{attallah2022ecg}} & \multirow{2}{*}{\makecell[l]{2-class classification \\(COVID-19 and normal)}} & ECG-BiCoNet & 98.80\% & 0.9880 \\
    \cline{3-5} 
       &   & \textbf{Ours} & \textbf{99.02\%} & \textbf{0.9901} \\
    \cline{2-5} 
       & \multirow{2}{*}{\makecell[l]{3-class classification\\(COVID-19, normal and abnormal)}} & ECG-BiCoNet & 91.73\% & 0.9150 \\
    \cline{3-5} 
       &   & \textbf{Ours} & \textbf{98.45\%} & \textbf{0.9793} \\
    \hline 
    
     
    \end{tabular}}
\end{table}
        
\section*{Discussions}
\label{section: discussions}
    In addition to having hazardous effects on human health, the COVID-19 pandemic has placed a significant strain on the healthcare system. To alleviate this problem, a rapid and straightforward method for COVID-19 diagnosis is required. The use of a deep learning methodology using 1D signals converted from ECG images to detect COVID-19 shows potential as a novel diagnostic procedure. 
    
    When compared against previous works on the same dataset, our method performed competitively. The following reasons may contribute to the proposed method's better performance: (i) white pixels contain no information, although they account for a major portion of the image; (ii) the amount of data required to train a 2D CNN is greater than that of 1D CNN~\cite{wu2018comparison} while the experimental dataset contains a modest number of samples (250 images for COVID-19 class), and (iii) it is thought that for ECG signal processing, a CNN with a larger kernel size and a lower stride value (compared to values often used in image processing) would be more effective in capturing the temporal correlation of adjacent signal points~\cite{cheng2021ecg}.
    
    This study contributes to the literature in numerous ways in this regard. These contributions can be summarized as follows: 
    \begin{itemize}
        \item Our work, together with previous research, demonstrated the capability to automatically distinguish COVID-19 patients by ECG signals with high sensitivity, cost-effectiveness, and non-invasiveness.
        \item To classify paper-based ECG data, a new and effective technique was developed, in which signals from a standard 12-lead ECG are extracted from the image and utilized as input to a 1D CNN model. This is the first time the ECG signal extracted from ECG printouts was used to detect COVID-19, to the best of our knowledge.
        \item The following benefits are provided by the image-to-signal conversion: (i) Less computational costs;
        (ii) More robust classification, regardless of the ECG report template and the resolution of ECG images; 
        (iii) Less storage costs by using 1D data instead of 2D;
        \item Our experiments confirmed the differences in ECG data of patients with COVID-19 and persons without any cardiac findings, as well as patients with other CVDs. The findings of the experiments might indicate the presence of COVID-19-induced cardiovascular changes.
    \end{itemize}
    
    
    There are some drawbacks to our proposed method. Firstly, similar to all current image-signal conversion algorithms, our proposed algorithm is sensitive to the shape of the ECG signal. The algorithm cannot differentiate the ECG signal in different leads if their amplitude is too high and combine together (signals with excessive amplitude are caused by noise or abnormal electrical activity of the heart due to CVDs). 
   Furthermore, although the proposed method has been tested with variety of experimental scenarios, it still has to be evaluated with different datasets. 
    In future works, we expect to integrate a COVID-19 diagnosis module into a mobile application, which will help medical practitioners by providing a quick and accurate approach to identifying COVID-19. Finally, although our method only produced the final diagnosis, an additional output that highlights the anomalies in the ECG~\cite{le2022lightx3ecg} of COVID-19 patients can also be helpful to cardiologists. Other potential approaches will be considered include training deep learning models with hierarchical disease dependencies~\cite{pham2021interpreting} or demographic data~\cite{le2022enhancing}. 
    
\section*{Conclusion}
\label{section:conclusion}
    In this work, a novel and effective method for automatically detecting COVID-19 utilizing ECG printouts is proposed. The proposed approach is based on an image-to-signal conversion algorithm, which extracts 1D ECG signal from scanned ECG printouts and feeds it into the 1D SEResNet18 model. We have run experiments on various scenarios and analyzed the findings. We also simulated the data set-up of prior research to develop classification models, then compared the performance of those models with their findings to evaluate the effectiveness of utilizing the extracted signal. The findings suggest that the ECG signal may be used to distinguish COVID-19 patients from other groups, and the proposed technique outperforms previous studies using the same dataset. Additionally, the capacity of the proposed method to distinguish COVID-19 ECG can serve as evidence that there are changes in ECG signals of COVID-19 patients. 
    The proposed ECG-based COVID-19 diagnose is easily extended to real-time systems and executed on mobile applications in our future works. In addition, we also intend to further interpret the model's decision and discover how COVID-19 patients' ECG changes compared to healthy people. As a result, it may benefit healthcare providers by offering a quick and accurate way to detect COVID-19, as well as reducing hospital costs by avoiding unnecessary hospital visits.

\section*{Acknowledgments}
This work was funded by Vingroup Joint Stock Company (Vingroup SJC), Vingroup, and supported by Vingroup Innovation Foundation (VINIF) under project code VINIF.2021.DA00128.

\bibliographystyle{unsrt}

\end{document}